\documentclass[twocolumn,pre,letterpaper]{revtex4-1}

\usepackage{dcolumn}
\usepackage{amsmath}

\usepackage{graphicx}
\usepackage[caption=false]{subfig}
\usepackage{bbm}

\newcommand{\dd}{\mathrm{d}}
\newcommand{\Ord}{\mathrm{O}}

\begin{document}

\title{Strongly-connected percolation on directed lattices}
\author{M. E. J. Newman}
\affiliation{Department of Physics and Center for the Study of Complex Systems,\\University of Michigan, Ann Arbor, MI 48109, USA}
\author{P. Grassberger}
\affiliation{Jülich Supercomputing Centre, FZ Jülich, D-52425 Jülich, Germany}
\author{R. M. Ziff}
\affiliation{Department of Chemical Engineering and Center for the Study of Complex Systems,\\University of Michigan, Ann Arbor, MI 48109, USA}

\begin{abstract}
We study percolation on lattices with directed bonds, focusing on the behavior of strongly-connected percolation clusters---clusters in which every site is reachable from every other along a directed path.  We consider the two-dimensional square lattice and various globally isotropic arrangements of the directions of the bonds.  Performing simulations using a range of algorithmic approaches, we calculate high-precision values for critical exponents, fractal dimensions, crossing probabilities, and percolation thresholds for bond percolation with each bond arrangement.  We find that the critical behavior is in a distinctly different universality class from that of traditional undirected percolation, but that all bond arrangements appear to fall in the same universality class.
\end{abstract}

\maketitle

\section{Introduction}
\label{sec:introduction}
Percolation models address the properties of lattices or networks where the bonds or sites are occupied independently at random with some probability~$p$, and focus primarily on the geometry and statistics of the contiguous clusters of occupied sites that form and grow as $p$ is increased~\cite{SA92}.  Percolation models have numerous applications, starting with the actual percolation phenomena for which they are named, such as the spread of fluids through rock~\cite{GG78,LSD81,Sahimi94}, but also including the modeling of granular materials~\cite{OT98,Tobochnik99}, polymers~\cite{BHP95}, composite materials~\cite{BFD92} and porous media~\cite{Machta91,MG95,Hassan22}, resistor networks~\cite{ARC85}, forest fires~\cite{Henley93}, epidemics~\cite{Newman02c}, robustness of the Internet and other networks~\cite{CEBH00,CNSW00}, biological evolution~\cite{RJ94}, and social influence~\cite{Solomon00}.

Percolation has traditionally been studied on undirected lattices but it can also be extended to directed ones where each bond has a direction, and the structure of the clusters becomes more involved in the directed setting.  The most straightforward case, often referred to simply as ``directed percolation,'' occurs when all bonds point in the same direction or directions---such as only up or to the right on a square lattice~\cite{Grassberger97,Hinrichsen00,AraujoEtAl14,ZiffGulariBarshad86,GrassbergerDeLaTorre79}.  In this paper we study the richer case of lattices where the bonds may point in any direction.

Two different definitions of clusters are commonly used for such lattices.  \textit{Weakly-connected clusters} (or weakly-connected components) are groups of sites connected by following occupied bonds in any direction.  Weakly-connected clusters are what one gets if one simply ignores the directions of the bonds and treats the system as a standard undirected percolation problem, so the statistics for this case are identical to ordinary percolation.

\begin{figure}[b]
\begin{center}
\includegraphics[width=5.0cm]{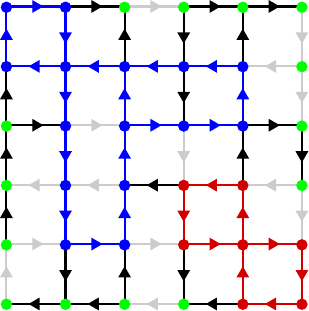}
\end{center}
\caption{Strongly-connected percolation on a directed square lattice.  In this example the directions of the bonds are chosen according to the ``Manhattan'' rule---see Section~\ref{sec:lattices}.  Bonds in gray are unoccupied; all others are occupied.  There are two large strongly-connected clusters, shown in blue and in red.  Each of the green sites is a single-site cluster on its own.  Note that two distinct clusters can be connected by occupied bonds, but only in one direction.  If they are connected in both directions they become the same cluster.}
\label{fig:scc}
\end{figure}

\textit{Strongly-connected clusters} (or components), on the other hand, are clusters of sites connected by occupied bonds such that every site in a cluster is reachable from every other along a directed path that follows bonds only in their forward direction---see Fig.~\ref{fig:scc}.  This means that every site in a strongly-connected cluster belongs to at least one closed cycle on the lattice (unless the cluster only has one site), since any pair of sites~$i,j$ in a cluster belong at a minimum to the cycle formed by the union of the path from $i$ to $j$ and the path from $j$ to~$i$.  Strongly-connected clusters have applications in transportation and traffic flow~\cite{VerbavatzBarthelemy21}, social network analysis~\cite{SDM16}, the world wide web~\cite{Broder00}, and bioinformatics~\cite{KMJ18,MZ03,Palmer25}, among other areas.  Strongly-connected clusters are similar to standard percolation clusters in covering the lattice completely and being non-overlapping---every site belongs to one, and only one, strongly-connected cluster---but in other respects they are quite different.  In particular, as we will see, their critical properties fall in a different universality class from ordinary percolation.

\begin{figure}
\begin{center}
\includegraphics[width=8.0cm]{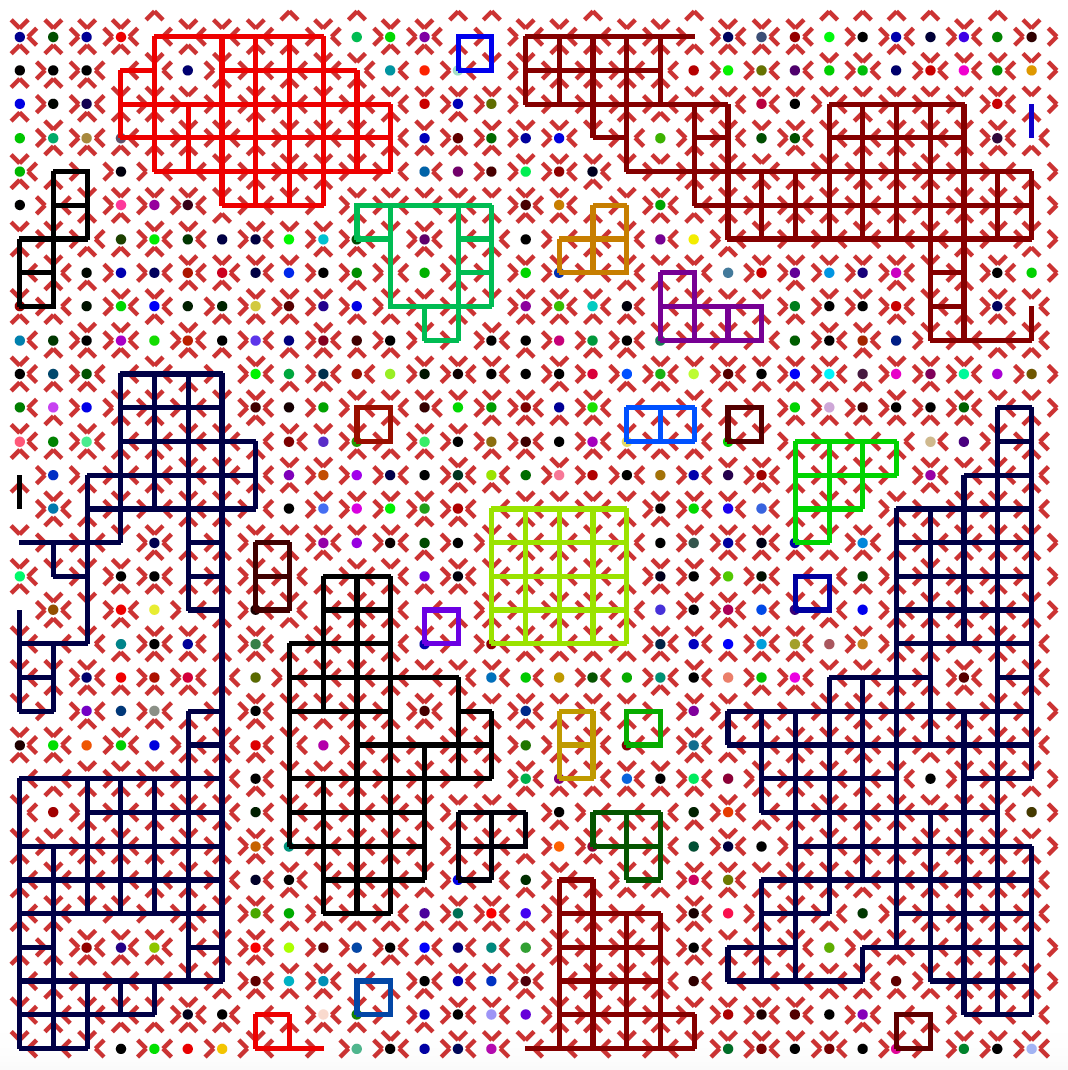}
\end{center}
\caption{Strongly-connected clusters for the random diode model on the square lattice at the percolation threshold $p_c = 1$.  Isolated sites are components of size~1.}
\label{fig:SCClattice}
\end{figure}

In this paper we focus on perhaps the simplest case of strongly-connected percolation, bond percolation in two dimensions on the square lattice.  In undirected percolation the system is completely specified once we fix the lattice, but in the directed case there are many ways the directed bonds can be arranged even on the same lattice.  Directions can be chosen randomly or according to a regular pattern.  Examples of random lattices are the random diode model, where each bond points randomly in one direction of the other with equal probability~\cite{Noronha18}, and the square ice model, where there are exactly two incoming and two outgoing bonds at each site but directions are otherwise random~\cite{Pauling35,Lieb67}.  An example of a non-random pattern is the so-called Manhattan lattice~\cite{Kasteleyn63,Barber70}, in which the vertical bonds alternate up and down and the horizontal ones alternate left and right.  Figure~\ref{fig:SCClattice} shows an example of strongly-connected clusters on a random diode lattice.

A somewhat different class of directed lattices, which we do not study here, are ones where some pairs of sites are connected by directed bonds in both directions, equivalent to an undirected bond.  Such lattices were studied by Redner~\cite{Redner81} and Dhar et al.~\cite{DharBarmaPhani81} in the form of the random resistor-diode model, in which every bond is randomly either a ``diode'' (it points in one direction) or a ``resistor'' (it points in both directions).  The percolation threshold on such lattices can be found by considering the outgoing bonds from each site, and satisfies
\begin{equation} 
\tfrac12 p_d + p_r = \tfrac12,
\label{eq:pdpc}
\end{equation}
where $p_d$ is the probability that a bond is a diode and is occupied, while $p_r$ is the probability that it is a resistor and is occupied.  De~Noronha et al.~\cite{Noronha18} have studied strongly-connected clusters for resistor-diode models on square, triangular, and hexagonal (honeycomb) lattices in two dimensions.  They measure various critical exponents, finding values that are similar for each of the lattices but different from those for traditional undirected percolation.  Recently, Wang and Li~\cite{WangLi26} have also studied strongly-connected clusters on resistor-diode lattices, in higher dimensions and on the complete graph and Costa et al.~\cite{CostaNoronhaVieiraAndradeMoreira25} studied them on the diamond and Wheatstone-bridge hierarchical lattices, calculating exact critical exponents and fractal dimensions for these systems.

The appearance of different critical exponent values suggests the existence of a new universality class for strongly-connected percolation.  In this paper we explore this possibility by examining the statistics of strongly-connected clusters on four different directed square lattices, focusing on systems where each bond points in a single direction only, by contrast with the resistor-diode configurations of previous studies.  The four configurations we study are the Manhattan lattice, the L-lattice, the random diode model, and the square ice model, which are defined in detail in Section~\ref{sec:lattices}.  These four configurations are not the only possibilities.  There are others of potential interest such as the two-neighbor model~\cite{Coupier24}, the biased directed percolation model \cite{ZhouYangZiffDeng12}, or the randomly-oriented Manhattan lattice~\cite{LedgerTothValko18}.  We conjecture that these should show strongly-connected percolation behavior similar to that of the lattices we study, presumably in the same universality class.

A secondary motivation for our work is the development of algorithms to efficiently generate and measure strongly-connected clusters on lattices or networks.  The standard algorithms of Kosaraju~\cite{Sharir81,CLRS01} and Tarjan~\cite{Tarjan72} can exhaustively find all clusters, and these provide a good starting point, but we also develop methods for efficient detection of percolation, search algorithms for critical points, and incremental algorithms for maintaining strongly-connected structures as bonds are added to the lattice one by one, in the spirit of the bond-by-bond approach of~\cite{NZ00,NZ01} for ordinary percolation.  Code for the algorithms is available for download from \verb|www.umich.edu/~mejn/scc.zip|.  See the appendices for details on how to use the code and additional information about the specific calculations in this paper.

\section{Directed lattices}
\label{sec:lattices}
\begin{figure}
\begin{center}
\subfloat[Manhattan lattice]{\includegraphics[width=3.5cm]{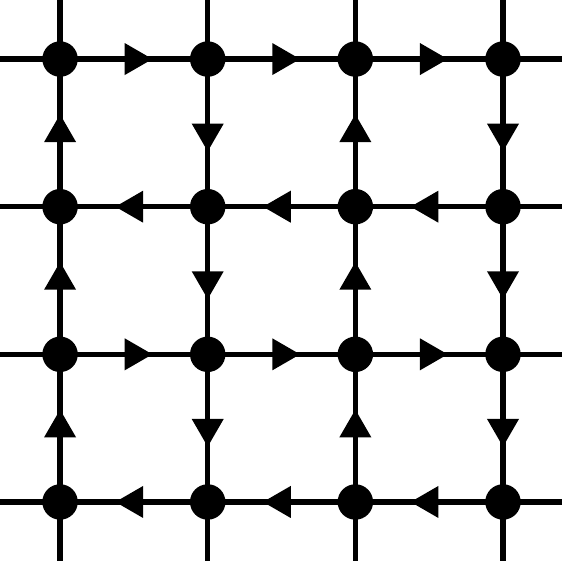}}\hspace{1cm}
\subfloat[L-lattice]{\includegraphics[width=3.5cm]{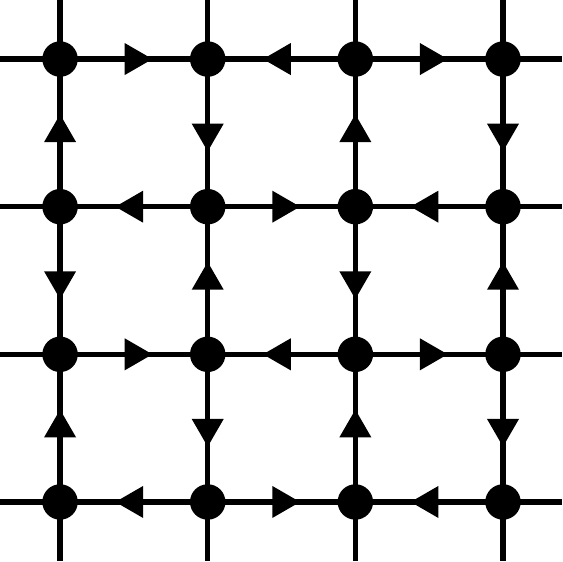}}\\[4ex]
\subfloat[Random diode model]{\includegraphics[width=3.5cm]{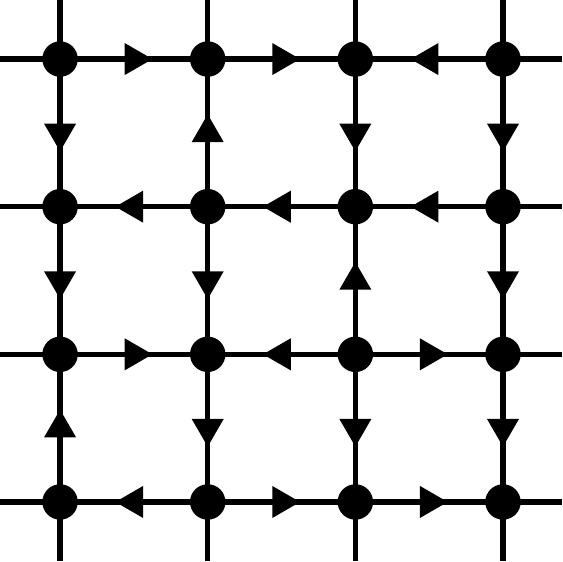}}\hspace{1cm}
\subfloat[Square ice model]{\includegraphics[width=3.5cm]{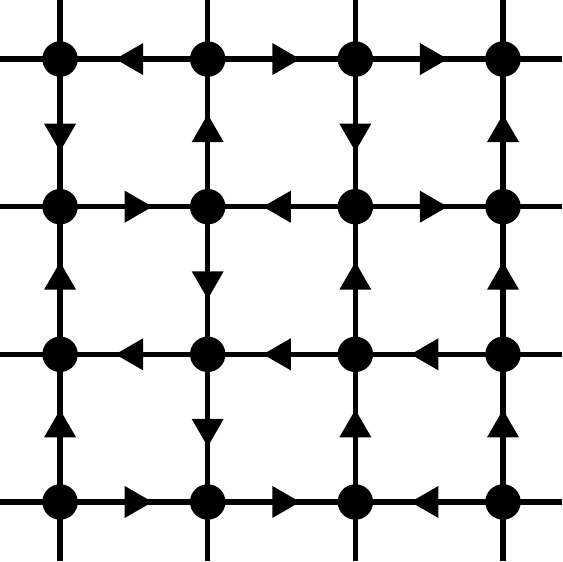}}
\end{center}
\caption{The four directed lattices that we study in this paper.}
\label{fig:models}
\end{figure}

We study square lattices in two dimensions with directed bonds and periodic boundary conditions, and we focus on four different arrangements of the directions of the bonds as shown in Fig.~\ref{fig:models}.

\paragraph{Manhattan lattice:} Named for the street patterns of Manhattan, New York, this lattice has east-west running ``streets'' that alternate directions, left then right, and similarly for north-south running streets (or ``avenues'' as they are called in Manhattan).  This means that every site has two ingoing and two outgoing bonds (Fig.~\ref{fig:models}a)~\cite{Kasteleyn63,Barber70}.

\paragraph{L-lattice:} On this lattice every site again has two ingoing and two outgoing bonds, but now the ingoing ones are opposite one another, and similarly for the outgoing ones (Fig.~\ref{fig:models}b)~\cite{Kasteleyn63,MannaGuttmann89}.  This means that every square plaquette forms a cycle (and hence a potential strongly-connected cluster, if all four bonds are occupied) and every directed path across the lattice must make a $90^\circ$ turn at every site it encounters (hence the name ``L-lattice''---the L refers to the $90^\circ$ turns).

\paragraph{Random diode model:} The remaining two configurations are randomized ones.  In the random diode model (Fig.~\ref{fig:models}c) the direction of each bond is chosen independently at random with 50-50 probability~\cite{Redner81}.  This means that it is possible, for example, to have four outgoing bonds at a single site, in which case that site is unreachable.

\paragraph{Square ice model:} In the square ice model, also called the six-vertex model, directions are again chosen randomly but with the constraint that each site must have two ingoing and two outgoing bonds \cite{Pauling35,Lieb67} (Fig.~\ref{fig:models}d).

For the randomized models, our percolation process involves two simultaneous forms of randomness: the randomness of the directions of the bonds and the inherent randomness of the percolation process itself, and measured quantities for these models are averages over both sources of randomness. The results do not depend (at least for finite-sized systems) on the order in which the averages are performed.

\begin{figure}
\begin{center}
\includegraphics[width=6.0cm]{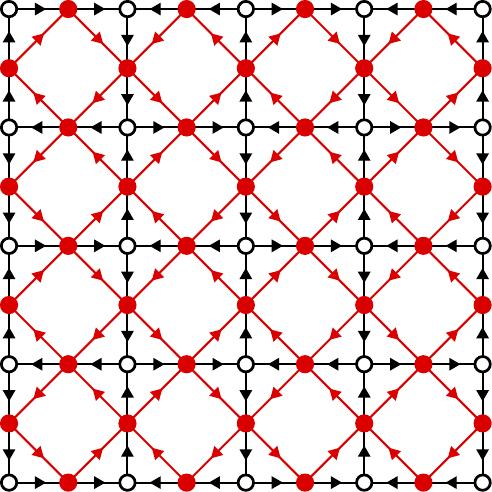}
\end{center}
\caption{Mapping from the L-lattice to the Manhattan lattice.  Bond percolation on the L-lattice (open circles, black) maps state-for-state onto site percolation on a Manhattan lattice at $45^\circ$ to the original (solid circles, red).}
\label{fig:mapping}
\end{figure}

Although our focus on this paper is on bond percolation, the methods we employ can easily be extended to site percolation as well.  We note that there is a direct formal equivalence between bond percolation on the L-lattice and site percolation on the Manhattan lattice, as illustrated in Fig.~\ref{fig:mapping}.  Starting with an L-lattice (open circles, in black), we place new sites at the centers of the bonds (solid circles, in red) and join them up to form a square lattice at $45^\circ$ as shown.  Sites on the new lattice are occupied if the corresponding bond was occupied on the original lattice and there exists a directed path between each pair of neighboring sites in the directions shown.  We note that the new lattice now takes the form of a Manhattan lattice and that there is a strongly-connected cluster on this lattice if and only if there existed such a cluster on the original lattice.  This immediately implies that all critical properties for site percolation on the Manhattan lattice are the same as those for bond percolation on the L-lattice, including critical exponents and scaling forms, as well as the position of the percolation threshold.  Hence we expect site percolation of strongly-connected clusters to be in the same universality class as bond percolation, as it is for ordinary percolation.  For other arrangements of bonds there is no equivalent simple mapping from bond to site percolation, but nonetheless we expect the universality class to remain the same.  (An equivalent mapping from site to bond percolation does exist for higher-dimensional analogs of the L-lattice, although we do not investigate these here.)

\section{Algorithms}
\label{sec:algorithms}
To probe the percolation properties of directed lattices numerically we use a combination of algorithmic approaches as follows.

\subsection{Breadth- and depth-first search algorithms}
\label{sec:dfs}
Under percolation on a directed lattice, the out-component of a site~$i$ is the set of sites reachable from~$i$ by following occupied bonds only in their forward direction, which can be found by standard breadth-first (or depth-first) search, running in time proportional to the number~$M$ of bonds on the lattice in the worst case.  One can also find the in-component of~$i$---the set of sites from which $i$ can be reached---by a reverse search that follows bonds only in their backward direction.  And the intersection of the in- and out-components of site~$i$ gives the strongly-connected cluster to which $i$ belongs---every site in the intersection can reach and is reachable from~$i$, and hence belongs to the same strongly-connected cluster.  If one wants to find only a single cluster, this is a good approach.  If one wants to find all clusters, it can be extended with some additional machinery, leading to Kosaraju's algorithm, whose operation centers around the construction of two overlapping depth-first searches~\cite{Sharir81,CLRS01}.

Kosaraju's algorithm is not, however, the most efficient method for finding all strongly-connected clusters.  A~more efficient method is the algorithm of Tarjan~\cite{Tarjan72}, which employs some additional bookkeeping strategies to find strongly-connected clusters with only a single depth-first search.  In this algorithm sites are labeled with an integer ``time'' index in the order in which they are discovered by the depth-first search, and each site~$i$ keeps a record of the earliest site (i.e.,~the smallest time index) reachable from~$i$, including itself.  At the end of the depth-first search the strongly-connected clusters are sets of sites that share a common earliest-reachable point.  Tarjan's algorithm is fast---like the standard algorithm for undirected clusters it runs in time~$\Ord(M)$.  This algorithm is the primary approach we use to find strongly-connected clusters in this paper.

\subsection{Incremental algorithms}
\label{sec:incremental}
In undirected percolation an alternative and competitive approach for finding clusters is the algorithm proposed in~\cite{NZ00}, in which (undirected) bonds are added one by one in random order to an initially empty lattice and one maintains a running tally of the percolation clusters as they form.  This can be done in an efficient manner by nominating one site to be the ``root'' of each cluster.  Every other site in the cluster has a pointer that either points to the root site, or to another site in the cluster such that by following a succession of pointers from site to site one can eventually reach the root.  This allows one to easily determine if two sites belong to the same cluster: their pointers will lead to the same root site if and only if they are in the same cluster.

This mechanism now allows us to efficiently update the cluster structure as bonds are added to the lattice.  When a new bond is added we find the roots for each of the two sites at its ends.  If the roots are different then the new edge has connected together two previously separate clusters and we update the lattice to reflect this by adding a pointer from the root of the smaller cluster to the root of the larger.  At the same time we perform the operation known as path compression: every time we traverse the pointers to reach the root of a cluster, we also change all pointers along the path traversed to point directly to the root.  This ensures that the next time we inquire about any site along the path, it will take only a single step to find the root.  It can be shown that this algorithm finds all clusters in time~$\Ord(M)$ on average, which is as fast as the standard breadth-first search~\cite{NZ01,Tarjan75}.

An advantage of this incremental assembly of the percolation clusters is that it gives us for free the cluster structure of every intermediate configuration through which the system passes.  If one starts with an empty lattice and adds every bond one by one in random order, then one gets one random configuration with every possible number of occupied bonds $m=0\ldots M$.  One can measure any property~$Q$ of those configurations, such as number of clusters, average or largest cluster size, or whether spanning has occurred, obtaining one value~$Q_m$ for each number of occupied bonds.  Then the expected value of $Q$ for any bond occupation probability~$p$ can be calculated from the binomial distribution of occupied bonds:
\begin{equation}
Q(p) = \sum_{m=0}^M {M\choose m} p^m (1-p)^{M-m} Q_m.
\label{eq:binomial}
\end{equation}
This allows one to calculate a complete curve of $Q(p)$ from $p=0$ to~1 in a single run that takes about the same time as the calculation for just one value of~$p$ using traditional breadth-first search.  For conventional percolation problems this approach has allowed researchers to perform a range of calculations that would have been impossible (or at least much harder) by conventional means~\cite{NZ00,NZ01,ZN02}.

Given this, it is natural to ask whether a similar approach could work for strongly-connected clusters on directed lattices.  The answer is yes, there are equivalent incremental algorithms for strongly-connected clusters, but they are less efficient than those for the undirected case.  The fundamental issue is that the addition of a single bond on a directed lattice does not necessarily join just two strongly-connected clusters.  Such a bond may complete a long cycle on the lattice that joins many smaller clusters into a single large one.  As a result the algorithms are more complex than for the undirected case.  One has to maintain a pointer structure for the clusters as before, but also an additional ``condensation graph,'' a network whose nodes represent the strongly-connected clusters and whose edges represent the directed connections between them.  The condensation graph is necessarily acyclic (contains no closed cycles), since if it contained any cycle then the nodes participating in that cycle would not be individual strongly-connected clusters (as the definition of the graph requires) but would be joined together into a single cluster.  As bonds are added to the lattice during the percolation process, any bond that completes a cycle on the condensation graph joins together the strongly-connected clusters along that cycle, which can be found by performing a depth-first search on the condensation graph.

The additional complexity affects the running time of the algorithm.  The standard version of the algorithm has a running time~$\Ord(M^{3/2})$~\cite{Haeupler12}, although recent advances using randomized methods have improved this slightly to~$\Ord(M^{4/3})$~\cite{BDP21}.  This is still worse than the corresponding algorithm for undirected percolation, and also worse than the linear running time of Tarjan's algorithm for strongly-connected clusters, although the authors of~\cite{Haeupler12} speculate that real-world running times may be better than the formal results suggest.  The incremental algorithm does provide complete information on percolation properties~$Q(p)$ over the entire range of $p$ from zero to one via Eq.~\eqref{eq:binomial}, which is useful for certain kinds of calculations, but Tarjan's algorithm has the edge in speed for calculations at just a single value of~$p$.

\subsection{Locating the percolation threshold}
\label{sec:binary}
An important application of the incremental approach in conventional percolation is finding the location of the percolation threshold.  Starting from an empty lattice one simply occupies bonds at random until percolation occurs.  One could do the same with the $\Ord(M^{4/3})$ algorithm of~\cite{BDP21} for strongly-connected percolation, but a faster and simpler approach is to combine the standard Tarjan algorithm with a binary search.  We first generate a random ordering of the bonds, then define $G(m)$ for $m=0\ldots M$ to be the lattice with the first $m$ bonds in that ordering occupied and let $m_c$ be the smallest value of $m$ such that $G(m)$ contains a percolating cluster.  (The specific criteria we use for identifying percolation are described in the next section.)

Suppose that we know $m_c$ lies in a window $[m_1,m_2]$.  For instance, we could initially set $m_1=0$ and $m_2=M$, which would guarantee that $m_c$ lies in the window.  Now we perform a binary search in this window.  We find the midpoint $m_\text{mid} = \bigl\lfloor \frac12 (m_1+m_2) \bigr\rfloor$, construct the lattice $G(m_\text{mid})$, run Tarjan's algorithm on it to find the strongly-connected clusters, and determine whether percolation has occurred.  If it has then we can narrow our search for~$m_c$ to the smaller half window $[m_1,m_\text{mid}]$.  If it has not we can narrow the search to the half window $[m_\text{mid},m_2]$.  Then we repeat the process, narrowing the window by a factor of two on every round until it is just a single bond wide, at which point we have found the percolation threshold.

The number of times we need to half the window size to reach this point is $\log_2 M$, and given that each run of Tarjan's algorithm takes time~$\Ord(M)$, the total running time for this approach is $\Ord(M \log M)$, which is better than the $\Ord(M^{4/3})$ of the incremental algorithm and only slightly worse than the $\Ord(M)$ running time for finding the percolation point in regular undirected percolation using the algorithm of~\cite{NZ00}.

\subsection{Detecting percolation}
\label{sec:wrapping}
There are various definitions of what it means to have a percolating cluster on a lattice.  The most general definition, applicable even for purely topological percolation problems, is the appearance of a so-called giant cluster whose size scales with the size of the lattice.  When the system is embedded in a space of finite dimension, a simpler definition is the appearance of a spanning cluster.  If the system has open boundaries this means a cluster that reaches from one side of the system to the other.  In this paper we study systems with periodic boundaries, for which the equivalent condition is the appearance of a wrapping cluster, one that wraps all the way around the boundary conditions and meets itself on the other side.

Efficient numerical simulation requires a way to rapidly detect such wrapping.  In this paper we again make use of the pointer-based algorithm of~\cite{NZ00} to perform this task, but modify the algorithm as described in~\cite{NZ01} so that each pointer does not simply identify its target site but becomes a vector that records the geometric displacement to the target.  By summing the vectors along the path from any site to the root of a cluster one can calculate the net displacement to the root.  When a bond is added to the lattice that connects two sites already in the same cluster then by definition the vectors to the common root of those two sites must differ by exactly the length of the bond.  The only exception is if the added bond wraps around the boundary conditions to join two sites that are an entire lattice size apart, in which case the vectors to the common root will differ by the lattice size minus the length of the bond.  By calculating the difference between the two vectors one can thus detect when wrapping has occurred.  Furthermore, on the square lattice one can distinguish between wrapping in the horizontal and vertical directions according to the vector direction of the difference, and hence one can separately find thresholds for horizontal and vertical wrapping.  In the square systems we study in this paper these thresholds are the same in expectation, but by measuring both we can get twice the statistics for essentially no extra work.  It is also possible for wrapping to occur in both directions, or for clusters to wrap multiple times around the boundary conditions in a spiraling fashion, and we can detect each of these conditions using the same vector difference method.

\subsection{Sampling configurations of the square ice model}
\label{sec:icealg}
For the particular case of the square ice model, we have an additional computational task to perform, of generating configurations of the lattice, which must be drawn uniformly from the set of all configurations in which every site has two ingoing and two outgoing bonds.  We do this by Markov chain Monte Carlo simulation.  The best known algorithm starts with any legal configuration and inverts the direction of a single bond, creating two defects at its ends, sites that have three ingoing or three outgoing bonds.  Then we invert further bonds along a chain from one of the defects, causing the defect to random walk around the lattice until, by chance, it encounters the other defect and the two annihilate, leaving a new perfect state of the lattice.  Repeating this process a sufficiently large number of times will then generate a new, independent configuration of the bonds.

This algorithm, however, is known to have slow mixing.  Here we employ the faster cluster-based algorithm of~\cite{BN98}, which makes use of the known equivalence between square ice and the zero-temperature three-state Potts antiferromagnet.  The latter admits a simple, global Monte Carlo update rule: we choose two of the three Potts spin values at random, construct all contiguous clusters made of those two values, and then for each cluster independently with 50-50 probability we either swap the values or not.  Previous studies of this algorithm indicate that it has a dynamic exponent not measurably different from zero, meaning that it takes a constant number of sweeps to generate new configurations of the lattice, regardless of system size.  Empirically, that number is about two sweeps~\cite{BN98}.  In our work, erring on the side of caution, we perform four sweeps for each new configuration we generate, ensuring that successive configurations are weakly correlated at best, then map the resulting Potts model back to the lattice of directed bonds before performing our percolation calculations.

\section{Results}
We focus on the behavior of strongly-connected clusters for bond percolation on the four directed lattices described in Section~\ref{sec:lattices}, and particularly on critical properties, including scaling behavior, the position of the percolation threshold, the wrapping probability at percolation, and critical exponents.

\subsection{Percolation threshold}
\label{sec:pc}
We employ a number of finite-size scaling approaches to estimate the critical bond occupation probability~$p_c$ on each of our lattices.  The first method is perhaps the most straightforward.  We measure the size of the out-component of a random site, i.e.,~the number of sites reachable along occupied bonds from that site including itself.  We expect such out-components to have the same critical behavior as percolation clusters in ordinary undirected percolation~\cite{Noronha18}.  This is most obvious for the random diode lattice, for which the out-components are precisely ordinary percolation clusters---see below.  For the other directed models there is no direct mapping to ordinary percolation, but the growth of an outgoing cluster is essentially an epidemic process in a random environment and hence we conjecture that it should belong to the ordinary percolation class.  Our numerical results support this conjecture.

Defining $P(s)$ to be the probability that an out-component has size greater than or equal to~$s$, we expect that in the critical region we should have a scaling law as a function of bond occupation probability~$p$ of the form
\begin{equation}
P(s) = s^{2-\tau} f\bigl((p-p_c) s^\sigma\bigr),
\label{eq:ordinary_scaling}
\end{equation}
for some universal scaling function~$f$, with the exponents taking their standard values for ordinary percolation, which are known exactly to be $\tau = \frac{187}{91}$ and $\sigma = \frac{36}{91}$.  Expanding in the small quantity $p-p_c$ then gives
\begin{equation}
s^{\tau-2} P(s) = A + B (p-p_c) s^\sigma + \ldots,
\end{equation}
so a plot of $s^{\tau-2} P(s)$ against $s^\sigma$ should give a straight line to leading order, and that line should become horizontal when $p=p_c$ for large~$s$.

As discussed in Section~\ref{sec:dfs}, out-components can be found by ordinary breadth-first search, starting from a random site and following only occupied bonds, only in their forward direction.  One of our lattices, the random diode lattice of Fig.~\ref{fig:models}c, presents a particularly simple case because each bond on this lattice points in each direction with equal probability independently, meaning there is exactly a probability $\frac12 p$ that during breadth-first search the bond is occupied and we follow it in its forward direction.  But this means that the out-component we find is just an ordinary percolation cluster on an undirected square lattice with bond occupation probability~$\frac12 p$.  We know that the percolation threshold for undirected percolation on such a lattice is exactly $\frac12$ by duality, and hence the threshold for directed percolation on the random diode lattice must satisfy $\frac12 p = \frac12$, so $p_c=1$.  (One can also recover the same result from Eq.~\eqref{eq:pdpc} by setting $p_r = 0$.)  So in this case there is no need to actually perform the breadth-first search to measure $p_c$: we already know the answer.

For the other three lattices we generate out-components by breadth-first search on a large $L\times L$ lattice with periodic boundary conditions (we use $L = 2^{14} = 16\,384$) and since we are interested in the probability $P(s)$ that the component size is greater than~$s$ we can safely terminate the breadth-first search if the size ever reaches whatever maximum value we are interested in (we set the limit at $2^{18} = 262\,144$).  Given the large size of the lattice, this essentially eliminates the possibility that the breadth-first search will ever encounter the perimeter of the lattice, and hence removes finite-size effects from the calculation, while at the same time substantially improving the running time because it eliminates the large amount of time needed to explore the largest out-components.

\begin{figure}
\begin{center}
\includegraphics[width=\columnwidth]{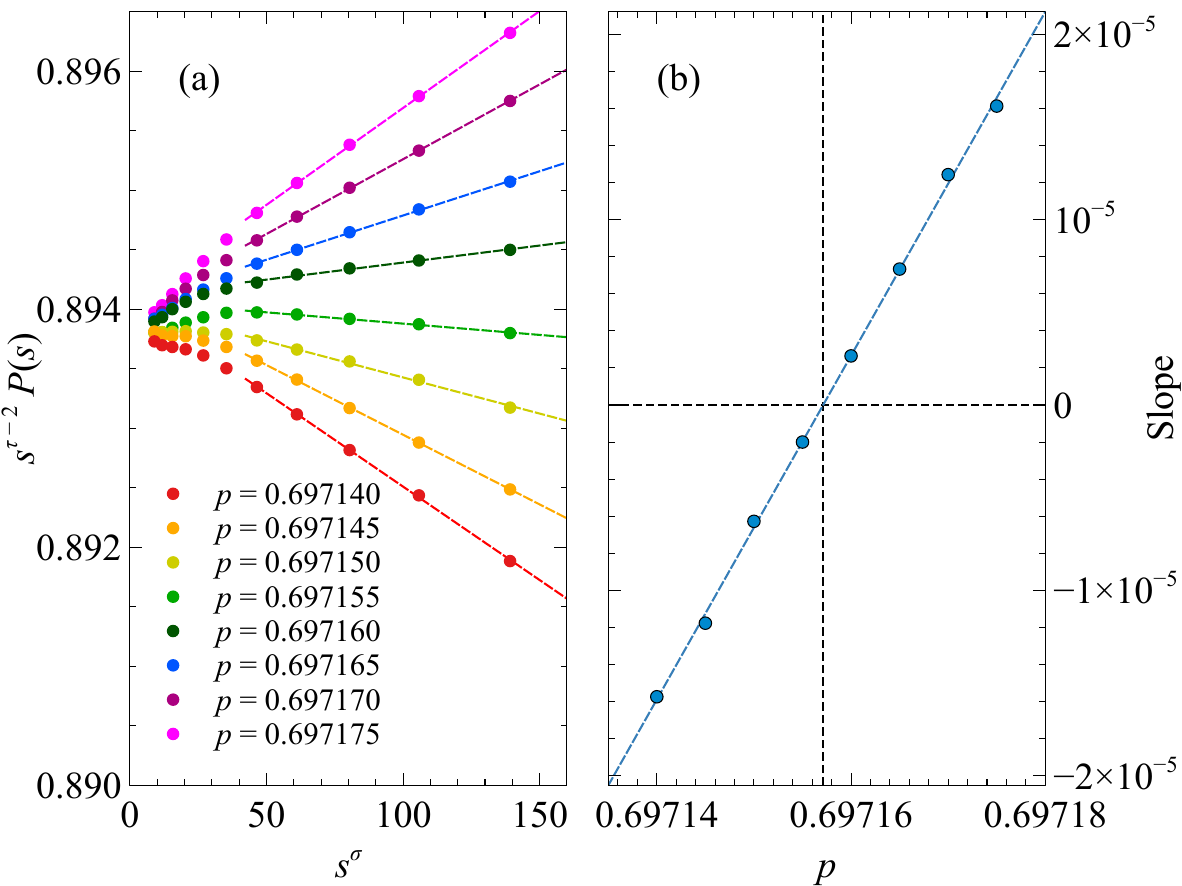}
\end{center}
\caption{(a)~The cumulative distribution of out-component sizes~$P(s)$ on the Manhattan lattice, plotted as $s^{\tau-2} P(s)$ against $s^\sigma$ as described in the text.  Dashed lines are straight-line fits to the rightmost five points in each case.  (b)~Slopes of the fits from panel~(a) as a function of~$p$.  The intercept where the slope crosses zero (vertical dashed line) is the best estimate of the position of the percolation threshold and gives $p_c = 0.6971571(5)$.}
\label{fig:mhpc}
\end{figure}

\begin{figure*}
\begin{center}
\includegraphics[width=15cm]{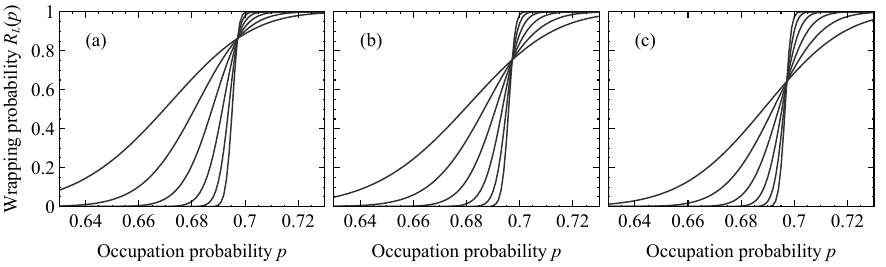}
\end{center}
\caption{The probability~$R_L(p)$ of a strongly-connected cluster wrapping around the periodic boundary conditions on $L\times L$ Manhattan lattices with $L=32$, 64, 128, 256, 512, and 1024 for each of three definitions of wrapping: (a)~wrapping in either direction, horizontally or vertically, (b)~wrapping in one particular direction (e.g.,~horizontally), and (c)~wrapping in both directions.}
\label{fig:mhrl}
\end{figure*}

From the results of these simulations we now calculate an estimate of~$P(s)$ and plot $s^{\tau-2} P(s)$ against $s^\sigma$ as discussed above.  Figure~\ref{fig:mhpc}a shows results for the Manhattan lattice for values of $p$ ranging from 0.697140 to 0.697175, and we can see that the data do indeed form (roughly) straight lines, which become horizontal around $p=0.697155$.  For a more accurate result we can extract the slope from fits at each value of~$p$ (dashed lines) and then make a plot of those slopes against~$p$ (Fig.~\ref{fig:mhpc}b).  Interpolating to estimate the point at which the slope crosses zero, we find that the percolation threshold falls at $p_c = 0.6971571(5)$ on the Manhattan lattice.  Similar calculations for our other lattices give $p_c=0.7401931(10)$ for the L-lattice, and 0.708834(1) for the ice model.  (The result for the L-lattice also implies that site percolation on the Manhattan lattice has the same threshold of $p_c=0.7401931(10)$, because of the mapping described in Section~\ref{sec:lattices}---see Fig.~\ref{fig:mapping}.)

One can perform equivalent calculations using strongly-connected clusters, but the results are less satisfactory for several reasons.  First, Tarjan's algorithm, while competitive with traditional breadth-first search in terms of speed, is more memory-hungry, requiring a significant amount of extra space for bookkeeping tasks, and this limits the size of the systems that we can study.  Second, the Tarjan algorithm can only identify a strongly-connected cluster once it is complete, which means one cannot save time by stopping the calculation when some predetermined size cutoff is reached.  And third, the strongly-connected clusters have different scaling exponents $\tau$ and~$\sigma$, whose values are known only approximately, which introduces an additional source of uncertainty into the calculations.  For this reason we favor the calculation based on out-components, which suffers from none of these issues.

An alternative approach for estimating $p_c$ is to look at the probability $R_L(p)$ at bond occupation probability~$p$ that there exists a strongly-connected cluster that wraps around the boundary conditions of a square $L\times L$ system, as discussed in Section~\ref{sec:wrapping}.  We consider three different definitions of wrapping: (a)~the first point at which wrapping occurs in either the horizontal or vertical direction, (b)~the first point at which wrapping occurs in a specified direction (e.g.,~horizontally), and (c)~the first point at which wrapping occurs in both directions.  (In these calculations we do not distinguish between wrapping once around the boundary conditions and wrapping multiple times---any wrapping is considered evidence of percolation.)

\begin{figure*}
\begin{center}
\includegraphics[width=15cm]{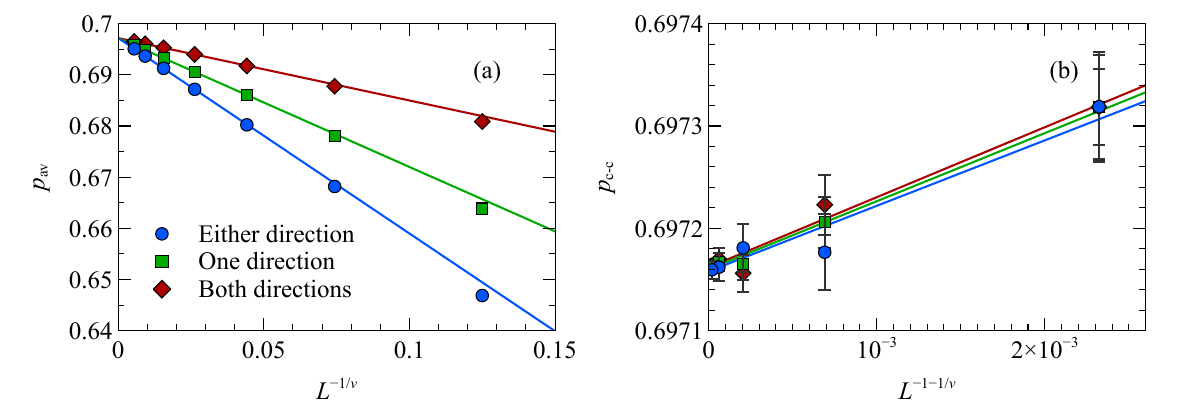}
\end{center}
\caption{Estimates of percolation thresholds on $L\times L$ Manhattan lattices of various sizes with at least a million repetitions of the percolation process for each size.  (a)~Scaling of the estimate~$p_\text{av}$, Eq.~\eqref{eq:pav}, as a function of $L^{-1/\nu}$ with $\nu = \frac43$ for values of $L$ from 16 to 1024 and for each of our three definitions of percolation: wrapping in either direction, horizontally or vertically (circles), wrapping in one particular direction, e.g.,~horizontally (squares), and wrapping in both directions (diamonds).  Error bars are smaller than the points in all cases.  Lines are fits to the left-most five points in each case.  (b)~Scaling of $p_\text{c-c}$, Eq.~\eqref{eq:c2c}, as a function of $L^{-1-1/\nu}$ for values of $L$ from 32 to 512, for the same three definitions of percolation.  Lines are fitted to all points.}
\label{fig:mhscaling}
\end{figure*}

The value of $R_L(p)$ can be computed by combining the binary search method of Section~\ref{sec:binary}, the pointer-based wrapping algorithm of Section~\ref{sec:wrapping}, and the binomial convolution of Eq.~\eqref{eq:binomial}.  Suppose we perform a total of $T$ runs of our binary search and find on run~$t$ that crossing first occurs when there $m_t$ occupied edges.  Then our estimate of $R_L(p)$ computed from an average over all runs~is
\begin{align}
R_L(p) &= \frac1T \sum_{t=1}^T \, \sum_{m=0}^M \binom{M}{m} p^m (1-p)^{M-m}
          \mathbbm{1}_{m\ge m_t},
\label{eq:RL}
\end{align}
where $\mathbbm{1}_x$ is the indicator function whose value is~1 when $x$ is true and 0 otherwise.  Figure~\ref{fig:mhrl} shows curves of $R_L(p)$ for each of the three types of wrapping on Manhattan lattices of various sizes, and similar curves can be calculated for the L-lattice and ice model (where in the latter case the curve is averaged over many configurations of the random bond directions).

\begin{table*}
\begin{tabular}{l|ccc|ccc|c}
        & \multicolumn{3}{c|}{Average probability at first wrapping~$p_\text{av}$} &
          \multicolumn{3}{c|}{Cell-to-cell RG estimate~$p_\text{c-c}$} \\
Lattice & Any direction & One direction & Both & Any direction & One direction & Both & Average \\
\hline
Manhattan & 0.697193(15) & 0.697187(12) & 0.697181(09) & 0.697158(4) & 0.697160(2) & 0.697162(4) & 0.697160(2) \\
L-lattice & 0.740227(11) & 0.740219(08) & 0.740211(05) & 0.740191(7) & 0.740194(7) & 0.740196(8) & 0.740193(4) \\
Ice model & 0.708852(02) & 0.708844(06) & 0.708839(06) & 0.708835(8) & 0.708839(5) & 0.708839(8) & 0.708838(4) \\
\end{tabular}
\caption{Scaling estimates of $p_c$ for the Manhattan, L-lattice, and ice models, using average and cell-to-cell estimates derived from three different definitions of wrapping around the boundaries on an $L\times L$ periodic system: wrapping in any direction (horizontal or vertical), wrapping in a specific direction, and wrapping in both directions.  Numbers in parentheses indicate standard errors on the trailing digits.}
\label{tab:pc}
\end{table*}

Given these curves we can compute a variety of estimates of~$p_c$~\cite{ZN02}.  Here we focus on two in particular.  The first is the average value of $p$ at which wrapping first occurs, which is given by
\begin{align}
p_\text{av} &= \int_0^1 p\,{\dd R_L\over\dd p} \>\dd p
   = 1 - \int_0^1 R_L(p) \>\dd p \nonumber\\
  &= 1 - {1\over T} \sum_{t=1}^T \sum_{m=0}^M {M\choose m} \mathbbm{1}_{m\ge m_t}
     \int_0^1 p^m (1-p)^{M-m} \>\dd p \nonumber\\
  &= {1\over T} \sum_{t=1}^T {m_t\over M+1},
\label{eq:pav}
\end{align}
where we have used integration by parts in the second equality, Eq.~\eqref{eq:RL} in the third, and $\int_0^1 p^m (1-p)^{M-m} \>\dd p = m!(M-m)!/(M+1)!$ in the fourth.
 
Our second measure of $p_c$ is the ``cell-to-cell'' renormalization group (RG) estimate~\cite{ZN02,ReynoldsStanleyKlein78}, which is the point at which $R_L(p)$ takes the same value for two different system sizes.  In our calculations we use system sizes~$L$ that are a power of two and define the cell-to-cell estimate~$p_\text{c-c}$ as the solution of
\begin{equation}
R_L(p) = R_{2L}(p).
\label{eq:c2c}
\end{equation}
In practice, we calculate the cell-to-cell estimates by computing $R_L(p)$ and $R_{2L}(p)$ from Eq.~\eqref{eq:RL} and then finding the root of $R_L(p)-R_{2L}(p)=0$ by binary search.

We expect $p_\text{av}$ to converge to $p_c$ with increasing system size as $p_\text{av}-p_c \sim L^{-1/\nu}$, where $\nu=\frac43$ is the standard correlation length exponent for percolation.  Figure~\ref{fig:mhscaling}a shows the results of measurements of $p_\text{av}$ against $L^{-1/\nu}$ for various sizes of Manhattan lattice for each of our three definitions of wrapping.  The vertical axis intercept gives us an estimate of $p_c$ in the limit $L\to\infty$ and we find that $p_c = 0.697193(15)$ for wrapping in any direction, 0.697187(12) for wrapping in one specific direction, and 0.697181(9) for wrapping in both directions.  Full results for all lattices are given in Table~\ref{tab:pc}.

As discussed in~\cite{ZN02}, the cell-to-cell RG estimate is expected to have a different and larger scaling exponent than~$p_\text{av}$, meaning it converges faster to the infinite size limit.  The value of the exponent is not known exactly, but we conjecture that $p_\text{c-c}-p_c \sim L^{-1-1/\nu} = L^{-1.75}$, which is in line with the behavior for ordinary percolation~\cite{ZN02} and gives a good fit to the data, as shown in Fig.~\ref{fig:mhscaling}b.  The exact value of the exponent is not crucial, however: we find that our estimates of $p_c$ do not depend strongly on it, within reason.  Values of $p_c$ for all lattices and definitions of wrapping are given in Table~\ref{tab:pc}.

Examining the table we note that the various values of $p_c$ from first wrapping are consistent with each other within the error bars, as are the values from the cell-to-cell measure, but that the first wrapping and cell-to-cell measures are about $2\sigma$ apart and hence arguably not fully consistent with each other.  We conjecture that this is due to corrections to scaling similar to those observed for ordinary percolation~\cite{ZN02}.  Such corrections would primarily affect~$p_\text{av}$, with its poorer leading-order scaling, which leads us to favor the cell-to-cell measure for estimating~$p_c$.  Combining the cell-to-cell estimates for the three different definitions of wrapping, we arrive at overall best estimates of $p_c = 0.697160(2)$ for the Manhattan lattice, 0.740193(4) for the L-lattice, and 0.708838(4) for the ice model, which are consistent with our values from measurements of out-component sizes at the start of this section.

\subsection{Crossing probability at threshold}
\label{sec:RL}
In addition to estimating~$p_c$, we can use the same simulation results in combination with Eq.~\eqref{eq:RL} to estimate the probability~$R_L(p)$ that a system has a wrapping cluster, as shown in Fig.~\ref{fig:mhrl}.  The value~$R_L(p_c)$ of this probability at the percolation threshold is of particular interest because we expect it to be universal: in the limit of large~$L$, and for any given definition of wrapping, it depends only on the large-scale cluster structure and is insensitive to small-scale detail such as the local arrangement of bonds on the lattice, and hence we expect that all arrangements---Manhattan, L-lattice, etc.---should have the same value of~$R_L(p_c)$.

\begin{figure}
\begin{center}
\includegraphics[width=\columnwidth]{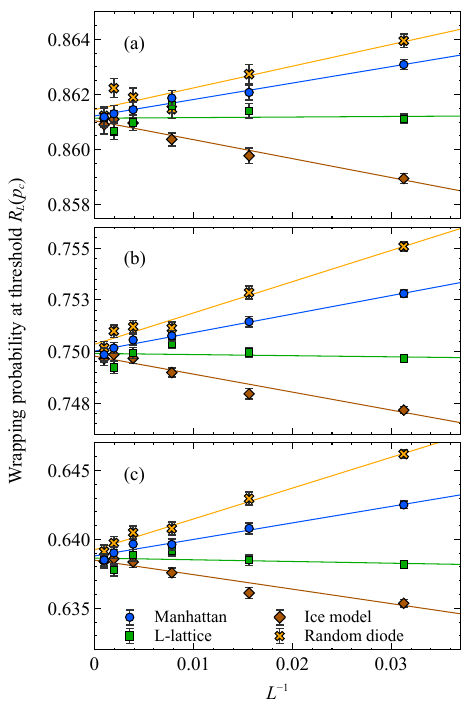}
\end{center}
\caption{Scaling of the wrapping probability at threshold~$R_L(p_c)$ for (a)~wrapping along any direction, (b)~wrapping along one specified direction (either horizontally or vertically), and (c)~wrapping along both directions.  Data points show values for the Manhattan lattice (blue circles), L-lattice (green squares), ice model (red diamonds), and random diode model (yellow crosses).}
\label{fig:pcross}
\end{figure}

Figure~\ref{fig:pcross} shows values of $R_L(p_c)$ calculated using our best estimates of~$p_c$ for the Manhattan lattice, L-lattice, and ice model as a function of~$L^{-1}$, while the straight-line fits give an estimate of the extrapolated values at $L=\infty$.  The extrapolated values are also given in Table~\ref{tab:pcross} and are consistent with the idea that $R_L(p_c)$ is universal.

\begin{table}
\setlength{\tabcolsep}{3.5pt}
\begin{tabular}{lccc}
Lattice & Any direction & One direction & Both \\
\hline
Manhattan & 0.86122(07) & 0.75001(09) & 0.63880(16) \\
L-lattice & 0.86114(21) & 0.74990(25) & 0.63864(31) \\
Ice model & 0.86105(11) & 0.74977(18) & 0.63849(25) \\
Random diode & 0.86145(26) & 0.75034(19) & 0.63924(17) \\
\hline
Average & 0.86118(06) & 0.75001(07) & 0.63889(10) \\
Ordinary perc. & 0.69047373 & 0.52105829 & 0.35164286 \\
\end{tabular}
\caption{Scaling estimates of the wrapping probability at the percolation threshold~$R_L(p_c)$ for the Manhattan lattice, L-lattice, ice model, and random diode model for each of the three definitions of wrapping used here.  Error bars include contributions due to uncertainty in the value of~$p_c$ as well as statistical error from the randomness of the percolation process.  We also give the best overall estimate of each probability based on a combination of the individual estimates and, for comparison, the corresponding probabilities for ordinary undirected percolation, which are known exactly~\cite{Pinson94,NZ01}.}
\label{tab:pcross}
\end{table}

Our fourth lattice, the random diode model, presents a special case because, as discussed in Section~\ref{sec:pc}, we know the position of the percolation threshold exactly---it falls at~$p_c=1$.  This means there is no need to measure the position of first wrapping or interpolate using Eq.~\eqref{eq:RL}.  We can simply run Tarjan's algorithm to find the strongly-connected clusters on a fully occupied lattice and then use the fast pointer-based algorithm of Section~\ref{sec:wrapping} to determine whether wrapping has taken place.  Figure~\ref{fig:pcross} also shows the results of these calculations for each of the three types of wrapping.  Extrapolation to $L=\infty$ gives estimates of $R_L(p_c)$ consistent with those from the other three lattices---see Table~\ref{tab:pcross} again.

Combining values across all four lattices, our best overall estimates of the three wrapping probabilities are 0.86117(6) for wrapping in any direction, 0.75001(7) for one specified direction, and 0.63867(11) for both directions.  These values differ considerably from the values for ordinary undirected percolation (which are listed in Table~\ref{tab:pcross} for comparison), clearly signaling that the two types of percolation lie in different universality classes.  We note that the probability for crossing in one direction across the square system is suspiciously close to 0.75.  The data are compatible---to within parts in~$10^5$---with the hypothesis that $R_L(p_c) = \frac34$ exactly for this definition of wrapping, although we know of no formal argument in favor of this value.

\begin{figure}
\begin{center}
\includegraphics[width=\columnwidth]{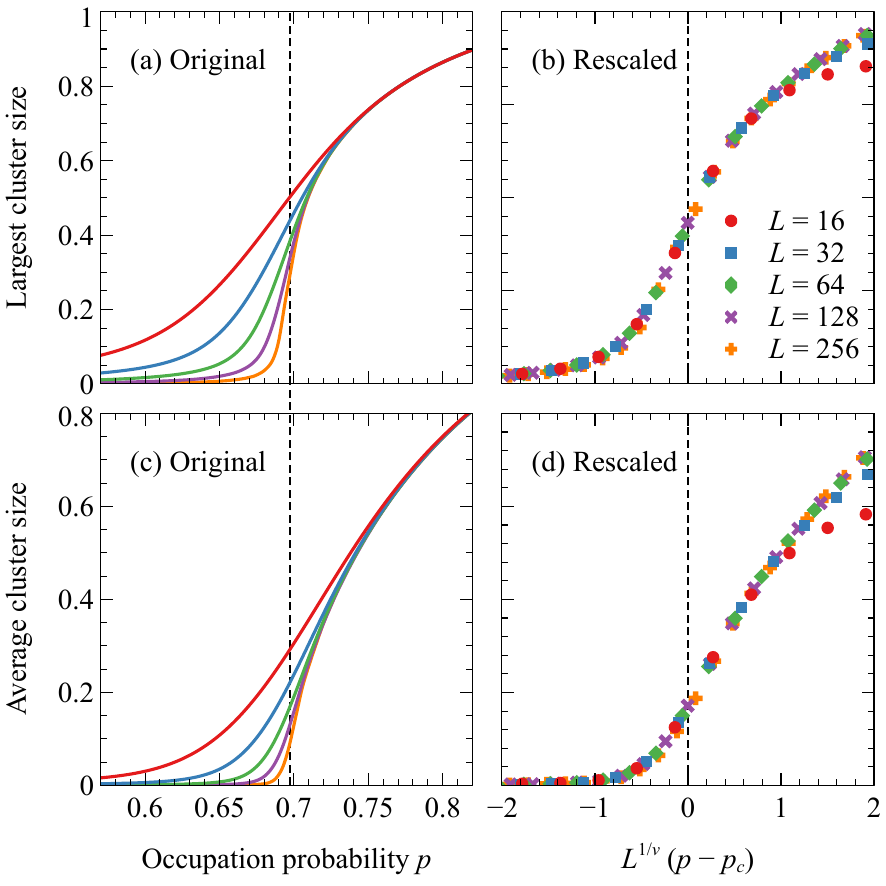}
\end{center}
\caption{(a)~The size~$S$ of the largest strongly-connected cluster divided by~$L^2$, as a function of bond occupation probability~$p$ on Manhattan lattices of size $L=16$, 32, 64, 128, and 256.  (b)~Scaling plot of $L^{\beta/\nu}S$ against $L^{1/\nu}(p-p_c)$.  Data are represented as individual points rather than lines to emphasize the collapse of multiple data sets onto a single curve.  (c)~The size $q/L^2$ of the average cluster to which a randomly chosen site belongs divided by~$L^2$, as a function of~$p$ on the same lattices.  (d)~Scaling plot of $L^{-\gamma/\nu}q$ against $L^{1/\nu}(p-p_c)$.  Curves were calculated using the incremental algorithm of Section~\ref{sec:incremental} combined with Eq.~\eqref{eq:binomial}.  The vertical dashed lines denote the position of the percolation transition on an infinite system.}
\label{fig:collapse}
\end{figure}

\subsection{Critical exponents}
\label{sec:exponents}
\begin{figure}
\begin{center}
\includegraphics[width=\columnwidth]{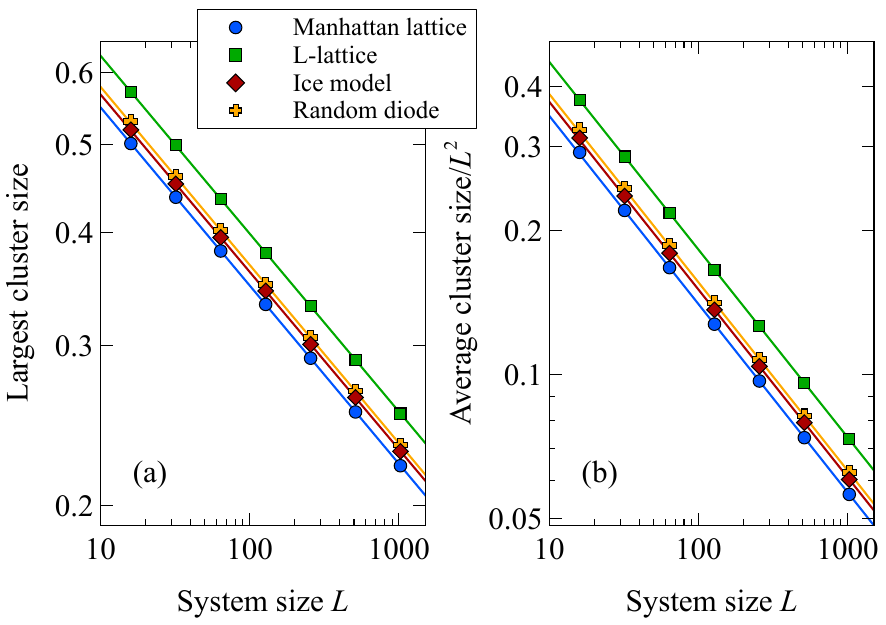}
\end{center}
\caption{(a)~Log-log plot of the size of the largest strongly-connected cluster at the percolation threshold as a function of system size~$L$ for each of the four lattices we study.  (b)~Average size of the cluster to which a random site belongs, divided by~$L^2$, as a function of~$L$.  Error bars are smaller than the data points in all cases.}
\label{fig:betagamma}
\end{figure}

Figure~\ref{fig:collapse}a shows a plot of the size~$S$ of the largest strongly-connected cluster on a Manhattan lattice as a fraction of the total size~$L^2$ of the system, calculated using the incremental algorithm described in Section~\ref{sec:incremental} and Eq.~\eqref{eq:binomial}.  The five curves are for system sizes $L=16$, 32, 64, 128, and 256 and in the critical region where the correlation length is large we expect these curves to obey a scaling law of the form
\begin{equation}
S = L^{-\beta/\nu} f\bigl(L^{1/\nu}(p-p_c)\bigr),
\label{eq:betascale}
\end{equation}
for some universal scaling function~$f$ and exponent~$\beta$.  Thus at the critical point $p=p_c$ we expect $S \sim L^{-\beta/\nu}$, and $\beta$ can be extracted by looking at the slope of a plot of $S$ against~$L$ on logarithmic scales.  Figure~\ref{fig:betagamma}a shows such a plot and provides a vivid demonstration of universality across our four lattices, since the slopes of the four lines are closely similar.  Based on fits to the right-most five points for each data set (which correspond to system sizes $L=64$, 128, 256, 512, and 1024) we estimate values of $\beta$ as shown in Table~\ref{tab:exponents}.  Combining these, our best overall estimate of the exponent is $\beta = 0.2618(3)$, distinctly different from the value $5/36 = 0.1389$ for ordinary undirected percolation.

\begin{table}
\setlength{\tabcolsep}{6pt}
\begin{tabular}{lcc}
Lattice & $\beta$ & $\gamma$ \\
\hline
Manhattan & 0.2623 & 2.1417 \\
L-lattice & 0.2618 & 2.1430 \\
Ice model & 0.2611 & 2.1429 \\
Random diode & 0.2621 & 2.1414 \\
\hline
Average & 0.2618(3) & 2.1423(4) \\
\end{tabular}
\caption{Estimates of the critical exponents $\beta$ and $\gamma$ from the slopes of the lines in Fig.~\ref{fig:betagamma}.}
\label{tab:exponents}
\end{table}

As a check on the scaling behavior, we can also make a plot of $L^{\beta/\nu}S$ against $L^{1/\nu}(p-p_c)$, which, following~\eqref{eq:betascale}, should give a picture of the scaling function~$f$ and hence result in a collapse of the data for different system sizes onto a single curve in the region close to the critical point.  Figure~\ref{fig:collapse}b shows such a scaling plot and a convincing collapse of the data.

The exponent $\gamma$ similarly governs the average cluster size~$q$, defined as the average size of the strongly-connected cluster to which a randomly chosen site belongs.  Figure~\ref{fig:collapse}c shows a plot of this quantity, again divided by~$L^2$ to bring the values into a smaller range for easy visualization.  Figure~\ref{fig:betagamma}b shows the corresponding log-log plot and the expected straight-line behavior $q/L^2 \sim L^{\gamma/\nu-2}$.  Fits to slopes of the lines in the figure (again using the right-most five points in each case) give estimates as shown in Table~\ref{tab:exponents}.  A combination of these figures gives a best overall estimate for the exponent of $\gamma = 2.1423(4)$, again distinctly different from the known value for undirected percolation, which is $\gamma = 43/18 = 2.389$.  Figure~\ref{fig:collapse}d shows the corresponding scaling collapse in a plot of $L^{-\gamma/\nu} q$ against $L^{1/\nu}(p-p_c)$.

\begin{figure}
\begin{center}
\includegraphics[width=\columnwidth]{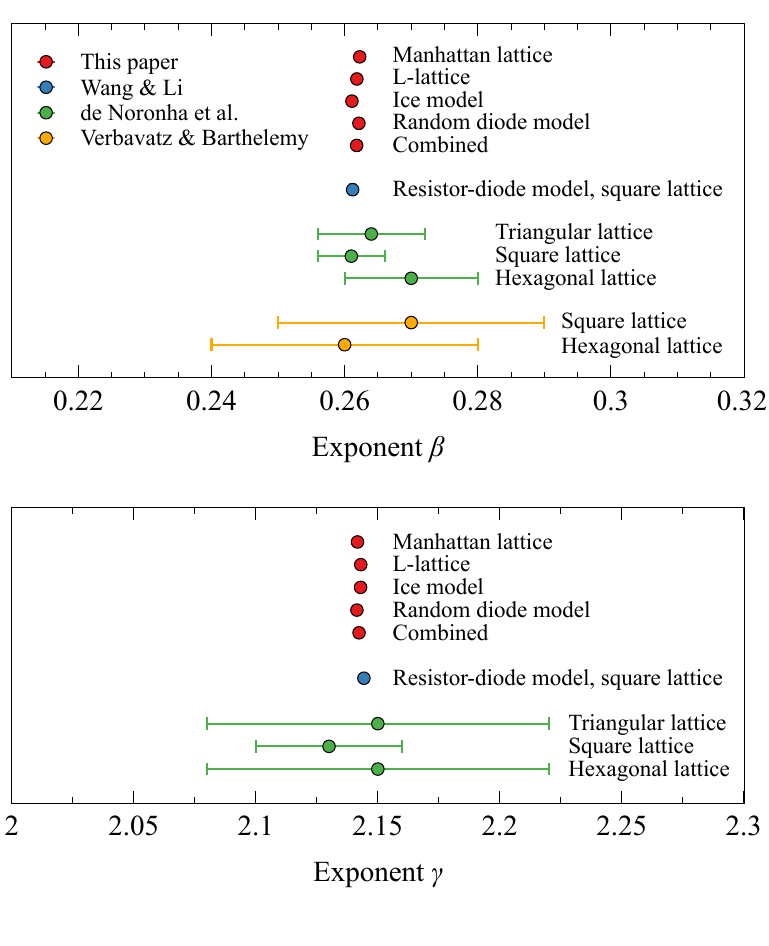}
\end{center}
\caption{Comparison of the values for the exponents $\beta$ and $\gamma$ found in this paper with those found previously for the random resistor-diode model on various lattices, as indicated, in work by de Noronha et al.~\cite{Noronha18}, Wang and Li~\cite{WangLi26}, and Verbavatz and Barthelemy~\cite{VerbavatzBarthelemy21}.  Error bars on the values from this paper and from Wang and Li~\cite{WangLi26} are plotted but are similar in size to the data points and hence mostly invisible.}
\label{fig:exponents}
\end{figure}

Figure~\ref{fig:exponents} compares our values for the critical exponents with those found for other models and lattices in previous work~\cite{Noronha18,WangLi26,VerbavatzBarthelemy21}, and agreement is good.

As a further test of our results, we note that $\beta$ and $\gamma$ should satisfy the hyperscaling relation
\begin{equation}
    \frac{2\beta}{\nu}+\frac{\gamma}{\nu} = d = 2.
\end{equation}
Using our best estimates of $\beta/\nu$ and $\gamma/\nu$ from the slopes in Fig.~\ref{fig:betagamma}, we find that $2\beta/\nu + \gamma/\nu = 1.9994(5)$, in excellent agreement with the expected behavior.

The fractal dimension~$d_f$ of the percolating strongly-connected cluster can be calculated from either $\beta$ or $\gamma$ as
\begin{align}
d_f &= d - {\beta\over\nu} =  1.8036(2) \nonumber\\
&= \frac12 \biggl(d+{\gamma\over\nu}\biggr) = 1.8033(2),
\label{eq:df}
\end{align}
with $d=2$.  These suggest a combined estimate of $d_f = 1.8035(1)$, which is is slightly lower than the value 1.8041(3) given recently by Wang and Li~\cite{WangLi26} but consistent with the values 1.801--1.805 given by de Noronha et al.\ \cite{Noronha18}.  Our value of $d_f$ also coincides with the exact value $1.80338003\ldots$ recently found by Costa et al.~\cite{CostaNoronhaVieiraAndradeMoreira25} for strongly connected clusters on the diamond hierarchical lattice.  This is a somewhat surprising result.  The diamond hierarchical lattice is two-dimensional but not periodic---it is a fractal lattice created by replacing each bond of an initial four-sided diamond with another diamond and repeating---so one would not necessarily assume it to fall in the same universality class as a periodic two-dimensional system.  Costa et al.\ also calculate the fractal dimension of ordinary (non-directed) percolation clusters on the same lattice and their value $d_f = 1.8928$ is nearly (though not exactly) equal to the known result $d_f = 91/48 = 1.8958333$ on periodic lattices.  Thus, the diamond hierarchical lattice appears to be a good, though not perfect, model for some properties of two-dimensional percolation.  We note that the other hierarchical lattice these authors study (the Wheatstone-bridge lattice, with effective dimension $\ln 5/\ln 2$) shows quite different behavior for $\beta$ and $d_f$.

Our results for $\beta$ and $d_f$ seem to rule out a conjecture of de Noronha et al.\ that $\beta$ takes exactly twice the value of the corresponding exponent for ordinary percolation $\beta = 2 \beta_\text{ordinary} = 5/18 = 0.2778$, which would imply that $d_f = 2-\beta/\nu = 43/24 = 1.7917$, assuming $\nu = \frac43$.  Costa et al.\ \cite{CostaNoronhaVieiraAndradeMoreira25} also note that this conjecture is violated by their results for the diamond hierarchical lattice.

In addition to the largest cluster size and average cluster size, we also examine the scaling of the complete distribution of cluster sizes.  Let $p(s)$ be the probability that a randomly chosen site belongs to a cluster of size~$s$.  In the critical region close to the percolation transition we expect this distribution to satisfy a scaling law of the form
\begin{equation}
    p(s) = s^{1-\tau} f\bigl( s/L^{d_f} \bigr)
    \label{eq:pscaling}
\end{equation}
where $\tau$ is the exponent governing strongly-connected clusters, which is in general different from the similarly named exponent for ordinary percolation that appears in Eq.~\eqref{eq:ordinary_scaling}.

\begin{figure}
\begin{center}
\includegraphics[width=\columnwidth]{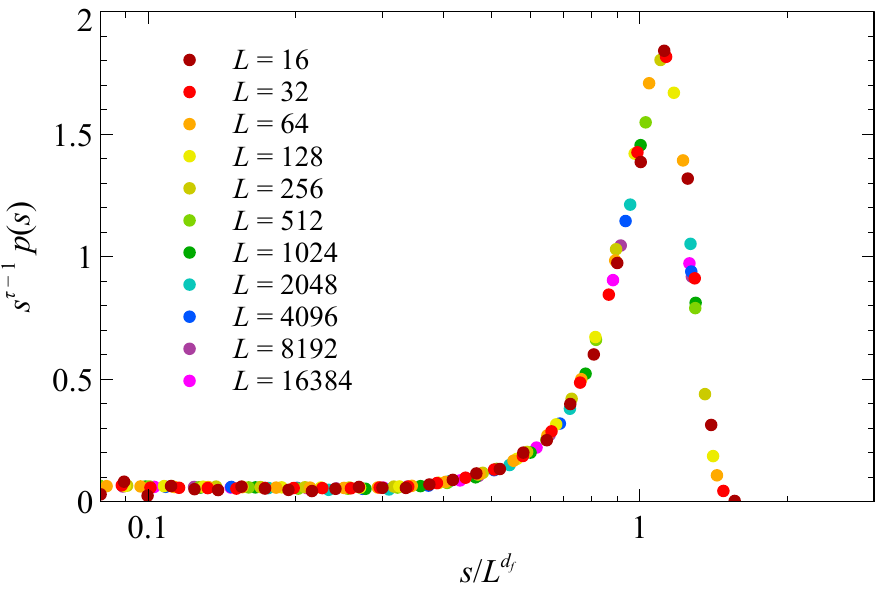}
\end{center}
\caption{Plot of $s^{\tau-1}p(s)$ vs.\ $s/L^{d_f}$ at the percolation threshold, for exponent values $\tau = 2.11$ and $d_f = 1.8035$, on Manhattan lattices of various sizes as indicated, showing a clear scaling collapse of the data.}
\label{fig:pscaling}
\end{figure}

Equation~\eqref{eq:pscaling} implies that a plot of $s^{\tau-1} p(s)$ as a function of $s/L^{d_f}$ should yield a collapse of the data for different system sizes onto a single curve.  We have an estimate of the value of $d_f = 1.8035$ from above and we can calculate $\tau$ from the scaling relation $\tau = 2 + \beta/(\beta+\gamma) \simeq 2.11$.  Figure~\ref{fig:pscaling} shows the resulting scaling plot, which gives a convincing collapse.

\subsection{Cluster hulls}
\label{sec:hull}
Hulls of clusters---their bounding perimeters, either outside or inside---have been studied in a variety of contexts in two-dimensional percolation~\cite{SapovalRossoGouyet85,ZiffCummingsStell84,SaleurDuplantier87,GrossmanAharony87,Kolb90,Manna89,Mandelbrot83,LawlerSchrammWerner03}.  The hull of a standard percolation cluster is known to have fractal dimension exactly~$\frac74$~\cite{SapovalRossoGouyet85,ZiffCummingsStell84,SaleurDuplantier87}, while the hull of the ``accessible perimeter" has dimension~$\frac43$~\cite{GrossmanAharony87,Kolb90}, as does the hull of the percolation backbone~\cite{Manna89}, as well as the Brownian hull~\cite{Mandelbrot83,LawlerSchrammWerner03} and many other two-dimensional objects.

\begin{figure}
\begin{center}
\includegraphics[width=4cm]{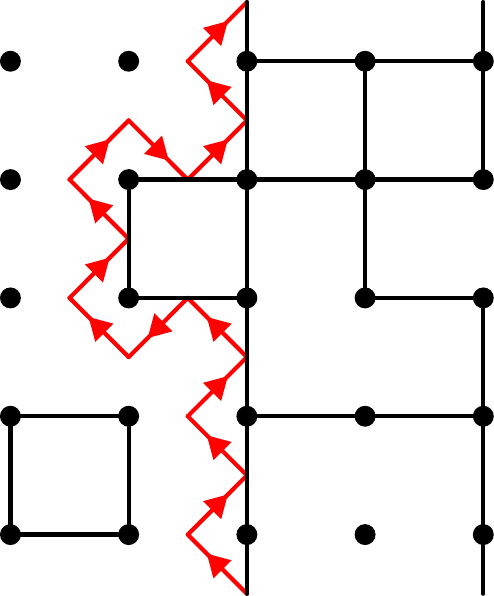}
\end{center}
\caption{A ``hull search walk'' that traces the external (or internal) hull of a percolation cluster by stepping diagonally between the centers of the bounding edges.}
\label{fig:hullwalk}
\end{figure}

\begin{figure}
\begin{center}
\includegraphics[width=6.5cm]{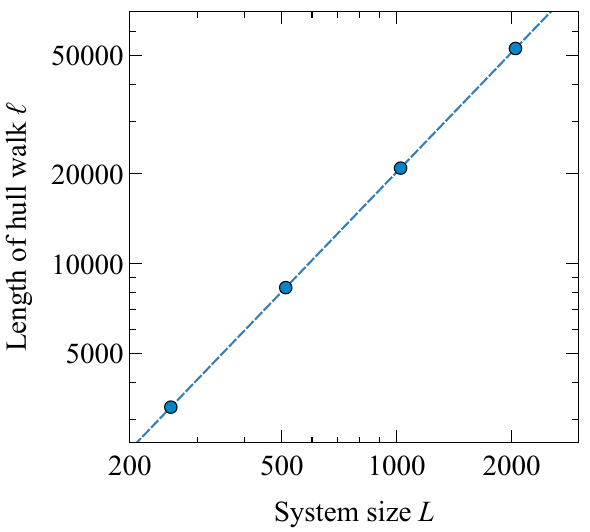}
\end{center}
\caption{Average length of a hull walk on a wrapping strongly connected cluster on the L-lattice for systems of various sizes.  The slope of the fit gives a value of $d_h = 1.3331(2)$ for the fractal dimension of the hull.  Error bars are smaller than the points.}
\label{fig:hulllength}
\end{figure}

To measure the hull of the clusters in strongly-connected percolation, we generate clusters on a square periodic L-lattice and identify the external hull of each cluster using an algorithm that starts from one occupied bond at the edge of the cluster and carries out a ``hull search walk,'' which steps along diagonals between the centers of occupied bonds and adjacent vacant bonds as shown in Fig.~\ref{fig:hullwalk}~\cite{GunnOrtuno85,Ziff89,MannaGuttmann89}.  Based on the resulting walk we first determine whether the cluster wraps around the periodic boundaries---when the walk returns to its starting point, if the net displacement in the $x$- or $y$-direction is a multiple of the lattice dimension~$L$, then wrapping has occurred.  Next, considering only the wrapping clusters, we determine the fractal dimension~$d_h$ of the hull from the variation of the average length of the hull walk~$\ell$ with system size: $\ell \sim L^{d_h}$.  Figure~\ref{fig:hulllength} shows a plot of $\ell$ against~$L$ on logarithmic scales, and the slope of the fit gives a value of $d_h = 1.3331(2)$, which is consistent with the hypothesis that $d_h = \frac43$ exactly, the same as the fractal dimension of the accessible perimeter in standard percolation.  An equivalent calculation on the random diode lattice also returns similar results.

This aligns with outward impressions---strongly-connected cluster hulls, like those in Fig.~\ref{fig:SCClattice}, appear more compact than the full perimeter of a normal percolation cluster (with $d_h = \frac74$) and similar to the accessible perimeter---but it would nonetheless be noteworthy if these two objects had exactly the same fractal dimension.  We know of no theoretical reason why this should be the case.

\section{Conclusions}
In this paper we have studied the percolation properties of strongly-connected clusters for bond percolation on directed lattices.  Focusing on the two-dimensional square lattice, we consider a variety of arrangements of the directions of the bonds, including both randomized arrangements such as the square ice model and non-random ones such as the Manhattan lattice.  Using a number of algorithmic approaches we measure various features of the strongly-connected clusters, with an emphasis on critical properties such as scaling behavior, position of the percolation threshold, critical exponents, and fractal dimensions.

We find convincing evidence that strongly-connected percolation falls in a different universality class from ordinary percolation in two dimensions, with distinctly different critical exponents, but that all arrangements of bond directions are in the same universality class.  This universality class also appears to be different from other known classes, such as those for standard (not strongly-connected) directed percolation (where $d_f \simeq 1.8405$ \cite{DengZiff22}), for the elastic backbone of ordinary percolation (where $d_f$ is known exactly $d_f = 1.6433\ldots$~\cite{NolinQianSunZhuang25}), or invasion percolation with trapping (where $d_f \simeq 1.82$~\cite{SheppardKnackstedtPinczewskiSahimi99}).

In addition to critical exponents we also measure the fractal dimension of the percolating cluster and of its bounding hull, and the probability of existence of a spanning or wrapping cluster at the percolation threshold.  Each of these quantities is expected to take a universal value.  The wrapping probability we find to be substantially higher than the corresponding probability for ordinary percolation, a perhaps surprising result, since the definition of a strongly-connected cluster is substantially more stringent than the definition of a cluster in ordinary percolation.

For the critical L-lattice and the random diode lattice we have also examined the external hull of wrapping strongly connected clusters, and find it to have a fractal dimension consistent with the exact value~$\frac43$, the same as the external (accessible) hull of regular percolation clusters~\cite{GrossmanAharony87}.  We conjecture that the same is true for external hulls in other isotropically directed systems.

The work described here could be extended in a number of directions.  One could look at other arrangements of the directed bonds on the square lattice, such as the two-neighbor model~\cite{Coupier24}, the biased directed percolation model~\cite{ZhouYangZiffDeng12}, the randomly-oriented Manhattan lattice~\cite{LedgerTothValko18}, or the biased Janus-particle directed model \cite{GarofaloAraujoEtAl26,HuZiffDeng22}.  Several researchers have looked previously at resistor-diode lattices in which occupied bonds can run either in one direction, as here, or in both directions~\cite{Redner81,DharBarmaPhani81,Noronha18,WangLi26}, focusing on the case of fully random resistors and diodes, but again other arrangements are possible, including both random and non-random variants.  One could also look at strongly-connected clusters under site (rather than bond) percolation on square lattices, or at either bond or site percolation on non-square lattices such as triangular or hexagonal lattices, or on cubic or hypercubic lattices in higher dimensions.  Some such cases have been investigated for the random resistor-diode model~\cite{Noronha18} but many other possibilities remain to be examined.

\begin{acknowledgments}
The authors thank Hao Hu for useful comments.  This work was funded in part by the US National Science Foundation under grant DMS--2404617.
\end{acknowledgments}

\appendix
\section{Computer code}
Code for algorithms used in this paper is available for download from \verb|www.umich.edu/~mejn/scc.zip| and is written in the C programming language for speed.  The code uses the GNU Scientific Library for the generation of random numbers and a copy of this library will be needed to compile and run the code.

The code is divided across a number of programs as follows:
\begin{itemize}
\item \texttt{binary.c}: Generates either a Manhattan lattice or an L-lattice on an $L\times L$ system of a specified size~$L$ and then repeatedly simulates percolation on that system, performing a binary search as described in Section~\ref{sec:binary} to locate the thresholds for wrapping of strongly-connected clusters in the horizontal and vertical directions.  Command-line parameters are the system size~$L$ and the number of repetitions.  The compile-time parameters \texttt{MANHATTAN} and \texttt{LLATTICE} in the code control which type of lattice is generated.  Output from the program has one line for each repetition of the calculation, consisting of the threshold number of occupied bonds at first wrapping in the horizontal and vertical directions.  Thus, a run of the program might look like this:
\begin{verbatim}
> binary 256 5
91417 91163
89188 90865
90751 90751
89824 90712
91541 91541
\end{verbatim}
In this instance, on the first of the five repetitions of the calculation the program has found that 91\,417 bonds had to be occupied before wrapping occurred in the horizontal direction, but only 91\,163 for wrapping in the vertical direction.
\item \texttt{clusters.c}: Generates either a Manhattan lattice or an L-lattice on an $L\times L$ system of a specified size~$L$ and then repeatedly simulates percolation on that lattice to measure the size of the largest strongly-connected cluster, the average cluster, and whether wrapping has occurred in the horizontal and vertical directions, at a specified value of the bond occupation probability~$p$.  Command-line parameters are the system size~$L$, the value of~$p$, and the number of repetitions.  The compile-time parameters \texttt{MANHATTAN} and \texttt{LLATTICE} in the code control which type of lattice is generated.  Output has one line for each repetition, consisting of the largest and average clusters sizes followed by two zero-one binary flags indicating the presence or absence of horizontal and vertical wrapping.  Thus a typical run might look like this:
\begin{verbatim}
> clusters 256 0.6971571 5
14117 3481.0364 1 0
13613 2989.0422 0 1
17806 5271.8834 1 0
19717 6119.7989 1 1
10670 2269.8085 0 1
\end{verbatim}
On the first of the five runs, in this case on a Manhattan lattice, the calculation has found a largest cluster size of 14\,117 and an average cluster size of 3481, and that wrapping has occurred in the horizontal direction but not the vertical direction.
\item \texttt{binary\_ice.c}: Generates random configurations of the ice model from Markov chain Monte Carlo simulations of the three-state Potts antiferromagnet as described in Section~\ref{sec:icealg}, then finds thresholds for wrapping in the horizontal and vertical directions by binary search.  Command-line parameters are the system size~$L$ and the number of repetitions of the calculation.  Output has one line per repetition, consisting of the threshold number of occupied bonds at first wrapping in the horizontal and vertical directions.
\item \texttt{clusters\_ice.c}: Generates random configurations of the ice model then measures the size of the largest strongly-connected cluster, the average cluster, and whether wrapping has occurred in the horizontal and vertical directions, at a specified value of the bond occupation probability~$p$.  Command-line parameters and output are the same as for \texttt{clusters.c}.
\item \texttt{clusters\_rl.c}: Generates configurations of the random diode model then measures the size of the largest strongly-connected cluster, the average cluster, and whether wrapping has occurred in the horizontal and vertical directions, at the percolation threshold~$p=1$.  Command-line parameters are the system size~$L$ and the number of runs.  Output is the same as for \texttt{clusters.c}.
\item \texttt{incremental.c}: Generates an $L\times L$ Manhattan or L-lattice system for given~$L$ and then repeatedly simulates percolation on it, finding the strongly-connected components using the incremental algorithm of Section~\ref{sec:incremental} and calculating the size of the largest component.  Command-line parameters are the system size~$L$ and the number of repetitions.  The compile-time parameters \texttt{MANHATTAN} and \texttt{LLATTICE} in the code control which type of lattice is generated.  Output is the largest cluster size divided by~$L^2$ for each number of occupied bonds from zero to~$2L^2$, averaged over all repetitions.
\end{itemize}

\section{Numerical calculations}
This appendix gives some further details of the numerical methods and calculations underlying the results in the main text.  For the results given in Table~\ref{tab:pc} and Figs.~\ref{fig:mhrl} and~\ref{fig:mhscaling} we performed at least a million simulations for each system size~$L$.  Estimates of $p_\text{av}$ and its uncertainty were made directly from Eq.~\eqref{eq:pav}.  Estimates of $p_\text{c-c}$ were made by computing~$R_L(p)$ and $R_{2L}(p)$ from Eq.~\eqref{eq:RL} and then solving for the root~$p$ of $R_{L}(p)-R_{2L}(p)=0$ by binary search.  The error on $p_\text{c-c}$ was estimated by bootstrapping with 100 bootstrap repetitions for each value of~$L$.  Final estimates of~$p_c$ are error-weighted straight-line extrapolations to the intercept at~$L=\infty$ and quoted errors are standard errors on the intercept.  Because the extrapolation assumes a straight line the calculation does not allow for corrections to scaling, as discussed in Section~\ref{sec:pc}.

The calculations of crossing probabilities in Section \ref{sec:RL} for Manhattan, L-lattice, and ice models are based on the same numerical results as the calculations of~$p_c$ in Section~\ref{sec:pc}, using Eq.~\eqref{eq:RL} to estimate the value of~$R_L(p_c)$, with best estimates of~$p_c$ for each lattice from Section~\ref{sec:pc}.  To allow for the uncertainty~$\sigma$ in the value of~$p_c$ we calculate $R_L(p)$ at $p_c\pm\sigma$ and estimate the resulting uncertainty in $R_L$ from the difference.  This uncertainty is then combined with the direct statistical uncertainty to estimate the total error on~$R_L$.  For the random diode model no such calculations are needed since we know the percolation threshold $p_c=1$ exactly (see Section~\ref{sec:pc}).  For this model we simply occupy all bonds, find all strongly-connected clusters directly, then test for wrapping using the algorithm of Section~\ref{sec:wrapping}.  Averaging over at least a million repetitions for each system size we then make a direct estimate of the probability of wrapping in each direction.

For the calculations of critical exponents in Section~\ref{sec:exponents} we perform direct measurements of the largest and average sizes of the strongly-connected clusters for each of our four lattices for system sizes up to~$L=1024$.  For the scaling collapses in Fig.~\ref{fig:collapse} we use the incremental algorithm of Section~\ref{sec:incremental}, which allows us to calculate complete curves of cluster size against~$p$.  For the results in Fig.~\ref{fig:betagamma} we perform simulations at $p_c$ and measure cluster sizes directly using Tarjan's algorithm (Section~\ref{sec:dfs}).  Again we perform at least a million repetitions for each system size, then fit the results as shown in Fig.~\ref{fig:betagamma} to extract the exponents.

For the calculation of the hull dimension~$d_h$ we generate individual strongly-connected clusters on the L-lattice by starting from a central site in an $L\times L$ system and finding the out-component and in-component of that site separately, then computing their intersection.  We generate at least ten million such clusters for each system size $L = 256$ to 2048 then identify the hull of each one using the hull search walk described in Section~\ref{sec:hull}.  Restricting ourselves to clusters that wrap in one direction, we then use the lengths of the walks to create the plot in Fig.~\ref{fig:hulllength} and fit the slope to extract the value of~$d_h$.


\begin{thebibliography}{10}
\expandafter\ifx\csname url\endcsname\relax
  \def\url#1{\texttt{#1}}\fi
\expandafter\ifx\csname urlprefix\endcsname\relax\def\urlprefix{URL }\fi

\bibitem{SA92}
D.~Stauffer and A.~Aharony, \textit{Introduction to Percolation Theory}. Taylor
  and Francis, London, 2nd edition (1992).

\bibitem{GG78}
P.~G. de~Gennes and E.~Guyon, Lois générales pour l'injection d'un fluide
  dans un milieu poreux aléatoire. \textit{J. de M\'ecanique} \textbf{3},
  403--432 (1978).

\bibitem{LSD81}
R.~G. Larson, L.~E. Scriven, and H.~T. Davis, Percolation theory of two phase
  flow in porous media. \textit{Chem. Eng. Sci.} \textbf{36}, 57--73 (1981).

\bibitem{Sahimi94}
M.~Sahimi, Long-range correlated percolation and flow and transport in
  heterogeneous porous media. \textit{J. Physique I} \textbf{4}, 1263--1288
  (1994).

\bibitem{OT98}
T.~Odagaki and S.~Toyofuku, Properties of percolation clusters in a model
  granular system in two dimensions. \textit{J. Phys. Cond. Mat.} \textbf{10},
  6447--6452 (1998).

\bibitem{Tobochnik99}
J.~Tobochnik, Granular collapse as a percolation transition. \textit{Phys. Rev.
  E} \textbf{60}, 7137--7142 (1999).

\bibitem{BHP95}
A.~Bunde, S.~Havlin, and M.~Porto, Are branched polymers in the universality
  class of percolation? \textit{Phys. Rev. Lett.} \textbf{74}, 2714--2716
  (1995).

\bibitem{BFD92}
S.~de~Bondt, L.~Froyen, and A.~Deruyttere, Electrical conductivity of
  composites: A percolation approach. \textit{J. Mater. Sci.} \textbf{27},
  1983--1988 (1992).

\bibitem{Machta91}
J.~Machta, Phase transition in fractal porous media. \textit{Phys. Rev. Lett.}
  \textbf{66}, 169--172 (1991).

\bibitem{MG95}
K.~Moon and S.~M. Girvin, Critical behavior of superfluid $^4${He} in aerogel.
  \textit{Phys. Rev. Lett.} \textbf{75}, 1328--1331 (1995).

\bibitem{Hassan22}
M.~K. Hassan, Recent development on fragmentation, aggregation and percolation.
  \textit{J. Phys. A} \textbf{55}, 191001 (2022).

\bibitem{ARC85}
L.~de~Arcangelis, S.~Redner, and A.~Coniglio, Anomalous voltage distribution of
  random resistor networks and a new model for the backbone at the percolation
  threshold. \textit{Phys. Rev. B} \textbf{31}, 4725--4728 (1985).

\bibitem{Henley93}
C.~L. Henley, Statics of a self-organized percolation model. \textit{Phys. Rev.
  Lett.} \textbf{71}, 2741--2744 (1993).

\bibitem{Newman02c}
M.~E.~J. Newman, Spread of epidemic disease on networks. \textit{Phys. Rev. E}
  \textbf{66}, 016128 (2002).

\bibitem{CEBH00}
R.~Cohen, K.~Erez, D.~{ben-Avraham}, and S.~Havlin, Resilience of the
  {I}nternet to random breakdowns. \textit{Phys. Rev. Lett.} \textbf{85},
  4626--4628 (2000).

\bibitem{CNSW00}
D.~S. Callaway, M.~E.~J. Newman, S.~H. Strogatz, and D.~J. Watts, Network
  robustness and fragility: Percolation on random graphs. \textit{Phys. Rev.
  Lett.} \textbf{85}, 5468--5471 (2000).

\bibitem{RJ94}
T.~S. Ray and N.~Jan, Anomalous approach to the self-organized critical state
  in a model for `life at the edge of chaos'. \textit{Phys. Rev. Lett.}
  \textbf{72}, 4045--4048 (1994).

\bibitem{Solomon00}
S.~Solomon, G.~Weisbuch, L.~de~Arcangelis, N.~Jan, and D.~Stauffer, Social
  percolation models. \textit{Physica A} \textbf{277}, 239--247 (2000).

\bibitem{Grassberger97}
P.~Grassberger, Directed percolation: Results and open problems. In S.~Puri and
  S.~Dattagupta (eds.), \textit{Nonlinearities in Complex Systems: Proceedings
  of the 1995 Shimla Conference on Complex Systems}, Narosa Publishing, New
  Delhi (1997).

\bibitem{Hinrichsen00}
H.~Hinrichsen, Non-equilibrium critical phenomena and phase transitions into
  absorbing states. \textit{Advances in Physics} \textbf{49}, 815--958 (2000).

\bibitem{AraujoEtAl14}
N.~Araújo, P.~Grassberger, B.~Kahng, K.~Schrenk, and R.~Ziff, Recent advances
  and open challenges in percolation. \textit{Eur. Phys. J.} \textbf{223},
  2307--2321 (2014).

\bibitem{ZiffGulariBarshad86}
R.~M. Ziff, E.~Gulari, and Y.~Barshad, Kinetic phase transitions in an
  irreversible surface-reaction model. \textit{Phys. Rev. Lett.} \textbf{56},
  2553--2556 (1986).

\bibitem{GrassbergerDeLaTorre79}
P.~Grassberger and A.~{de la Torre}, Reggeon field theory ({S}chlögl's first
  model) on a lattice: {M}onte {C}arlo calculations of critical behaviour.
  \textit{Annals of Physics} \textbf{122}, 373--396 (1979).

\bibitem{VerbavatzBarthelemy21}
V.~Verbavatz and M.~Barthelemy, From one-way streets to percolation on random
  mixed graphs. \textit{Phys. Rev. E} \textbf{103}, 042313 (2021).

\bibitem{SDM16}
S.~Dhingra, P.~S. Dodwad, and M.~Madan, Finding strongly connected components
  in a social network graph. \textit{Int. J. Computer Applications}
  \textbf{136}, 7 (2016).

\bibitem{Broder00}
A.~Broder, R.~Kumar, F.~Maghoul, P.~Raghavan, S.~Rajagopalan, R.~Stata,
  A.~Tomkins, and J.~Wiener, Graph structure in the web. \textit{Computer
  Networks} \textbf{33}, 309--320 (2000).

\bibitem{KMJ18}
S.~Kumar, S.~Mahajan, and S.~Jain, Feedbacks from the metabolic network to the
  genetic network reveal regulatory modules in {E.}\ coli and {B.}\ subtilis.
  \textit{PLOS One} \textbf{13}, e0203311 (2018).

\bibitem{MZ03}
H.-W. Ma and A.-P. Zeng, The connectivity structure, giant strong component and
  centrality of metabolic networks. \textit{Bioinformatics} \textbf{19},
  1423--1430 (2003).

\bibitem{Palmer25}
P.~Palmer-Rodríguez, R.~Alberich, M.~Reyes-Prieto, J.~A. Castro, and
  M.~Llabrés, Metadag: A web tool to generate and analyse metabolic networks.
  \textit{BMC Bioinformatics} \textbf{26}, 31 (2025).

\bibitem{Noronha18}
A.~W.~T. de~Noronha, A.~A. Moreira, A.~P. Vieira, H.~J. Herrmann, J.~S.
  Andrade~Jr., and H.~A. Carmona, Percolation on an isotropically directed
  lattice. \textit{Phys. Rev. E} \textbf{98}, 062116 (2018).

\bibitem{Pauling35}
L.~Pauling, The structure and entropy of ice and of other crystals with some
  randomness of atomic arrangement. \textit{J. Am. Chem. Soc.} \textbf{57},
  2680--2684 (1935).

\bibitem{Lieb67}
E.~H. Lieb, Exact solution of the problem of the entropy of two-dimensional
  ice. \textit{Phys. Rev. Lett.} \textbf{18}, 692--694 (1967).

\bibitem{Kasteleyn63}
P.~Kasteleyn, A soluble self-avoiding walk problem. \textit{Physica}
  \textbf{29}, 1329--1337 (1963).

\bibitem{Barber70}
M.~Barber, Asymptotic results for self-avoiding walks on a {M}anhattan lattice.
  \textit{Physica} \textbf{48}, 237--241 (1970).

\bibitem{Redner81}
S.~Redner, Percolation and conduction in a random resistor-diode network.
  \textit{J. Phys. A} \textbf{14}, L349 (1981).

\bibitem{DharBarmaPhani81}
D.~Dhar, M.~Barma, and M.~K. Phani, Duality transformations for two-dimensional
  directed percolation and resistance problems. \textit{Phys. Rev. Lett.}
  \textbf{47}, 1238--1241 (1981).

\bibitem{WangLi26}
Q.~Wang and M.~Li, Percolation transition of strongly connected clusters in
  finite dimensions and on complete graphs. Preprint arxiv:2605.16987 (2026).

\bibitem{CostaNoronhaVieiraAndradeMoreira25}
S.~{\relax dos S}. Costa, A.~W.~T. de~Noronha, A.~P. Vieira, J.~S. Andrade~Jr.,
  and A.~A. Moreira, Critical exponents for isotropically directed percolation
  on hierarchical lattices. \textit{Phys. Rev. E} \textbf{111}, 054129 (2025).

\bibitem{Coupier24}
D.~Coupier, B.~Henry, B.~Jahnel, and J.~Köppl, The planar lattice two-neighbor
  graph percolates. Preprint arxiv:2412.20781 (2024).

\bibitem{ZhouYangZiffDeng12}
Z.~Zhou, J.~Yang, R.~M. Ziff, and Y.~Deng, Crossover from isotropic to directed
  percolation. \textit{Phys. Rev. E} \textbf{86}, 021102 (2012).

\bibitem{LedgerTothValko18}
S.~Ledger, B.~Tóth, and B.~Valkó, Random walk on the randomly-oriented
  {M}anhattan lattice. \textit{Electron. Commun. Probab.} \textbf{23}, 1--11
  (2018).

\bibitem{Sharir81}
M.~Sharir, A strong-connectivity algorithm and its applciations in data flow
  analysis. \textit{Computers and Mathematics with Applications} \textbf{7},
  67--72 (1981).

\bibitem{CLRS01}
T.~H. Cormen, C.~E. Leiserson, R.~L. Rivest, and C.~Stein, \textit{Introduction
  to Algorithms}. MIT Press, Cambridge, MA, 2nd edition (2001).

\bibitem{Tarjan72}
R.~E. Tarjan, Depth-first search and linear graph algorithms. \textit{SIAM J.
  Comput.} \textbf{1}, 146--160 (1972).

\bibitem{NZ00}
M.~E.~J. Newman and R.~M. Ziff, Efficient {M}onte {C}arlo algorithm and
  high-precision results for percolation. \textit{Phys. Rev. Lett.}
  \textbf{85}, 4104--4107 (2000).

\bibitem{NZ01}
M.~E.~J. Newman and R.~M. Ziff, Fast {M}onte {C}arlo algorithm for site or bond
  percolation. \textit{Phys. Rev. E} \textbf{64}, 016706 (2001).

\bibitem{MannaGuttmann89}
S.~S. Manna and A.~J. Guttmann, Kinetic growth walks and trails on oriented
  square lattices: Hull percolation and percolation hulls. \textit{J. Phys. A}
  \textbf{22}, 3113 (1989).

\bibitem{Tarjan75}
R.~E. Tarjan, Efficiency of a good but not linear set union algorithm.
  \textit{J. ACM} \textbf{22}, 215--225 (1975).

\bibitem{ZN02}
R.~M. Ziff and M.~E.~J. Newman, Convergence of threshold estimates for
  two-dimensional percolation. \textit{Phys. Rev. E} \textbf{66}, 016129
  (2002).

\bibitem{Haeupler12}
B.~Haeupler, T.~Kavitha, R.~Mathew, S.~Sen, and R.~E. Tarjan, Incremental cycle
  detection, topological ordering, and strong component maintenance.
  \textit{ACM Transactions on Algorithms} \textbf{8}, 3 (2012).

\bibitem{BDP21}
A.~Bernstein, A.~Dudeja, and S.~Pettie, Incremental {SCC} maintenance in sparse
  graphs. In \textit{Proceedings of the 29th Annual European Symposium on
  Algorithms (ESA 2021)}, p.~14, Dagstuhl Publishing, Germany (2021).

\bibitem{BN98}
G.~T. Barkema and M.~E.~J. Newman, {M}onte {C}arlo simulation of ice models.
  \textit{Phys. Rev. E} \textbf{57}, 1155--1166 (1998).

\bibitem{ReynoldsStanleyKlein78}
P.~J. Reynolds, H.~E. Stanley, and W.~Klein, Percolation by position-space
  renormalisation group with large cells. \textit{J. Phys. A} \textbf{11}, L199
  (1978).

\bibitem{Pinson94}
H.~T. Pinson, Critical percolation on the torus. \textit{J. Stat. Phys.}
  \textbf{75}, 1167--1177 (1994).

\bibitem{SapovalRossoGouyet85}
B.~Sapoval, M.~Rosso, and J.~Gouyet, The fractal nature of a diffusion front
  and the relation to percolation. \textit{J. Physique Lett.} \textbf{46},
  149--156 (1985).

\bibitem{ZiffCummingsStell84}
R.~M. Ziff, P.~T. Cummings, and G.~Stell, Generation of percolation cluster
  perimeters by a random walk. \textit{J. Phys. A} \textbf{17}, 3009 (1984).

\bibitem{SaleurDuplantier87}
H.~Saleur and B.~Duplantier, Exact determination of the percolation hull
  exponent in two dimensions. \textit{Phys. Rev. Lett.} \textbf{58}, 2325--2328
  (1987).

\bibitem{GrossmanAharony87}
T.~Grossman and A.~Aharony, Accessible external perimeters of percolation
  clusters. \textit{J. Phys. A} \textbf{20}, L1193 (1987).

\bibitem{Kolb90}
M.~Kolb, Crossover from standard to reduced hull for random percolation.
  \textit{Phys. Rev. A} \textbf{41}, 5725--5727(R) (1990).

\bibitem{Manna89}
S.~S. Manna, Structure of backbone perimeters of percolation clusters.
  \textit{J. Phys. A} \textbf{22}, 433 (1989).

\bibitem{Mandelbrot83}
B.~B. Mandelbrot, \textit{The Fractal Geometry of Nature}. W. H. Freeman, New
  York (1983).

\bibitem{LawlerSchrammWerner03}
G.~Lawler, O.~Schramm, and W.~Werner, Test of scaling exponents for
  percolation-cluster perimeters. \textit{J. Amer. Math. Soc.} \textbf{16},
  917--955 (2003).

\bibitem{GunnOrtuno85}
J.~M.~F. Gunn and M.~Ortuño, Percolation and motion in a simple random
  environment. \textit{J. Phys. A} \textbf{18}, L1095 (1985).

\bibitem{Ziff89}
R.~M. Ziff, Hull-generating walks. \textit{Physica D} \textbf{38}, 377--383
  (1989).

\bibitem{DengZiff22}
Y.~Deng and R.~M. Ziff, The elastic and directed percolation backbone.
  \textit{J. Phys. A} \textbf{55}, 244002 (2022).

\bibitem{NolinQianSunZhuang25}
P.~Nolin, W.~Qian, X.~Sun, and Z.~Zhuang, Backbone exponent and annulus
  crossing probability for planar percolation. \textit{Phys. Rev. Lett.}
  \textbf{134}, 117101 (2025).

\bibitem{SheppardKnackstedtPinczewskiSahimi99}
A.~P. Sheppard, M.~A. Knackstedt, W.~V. Pinczewski, and M.~Sahimi, Invasion
  percolation: new algorithms and universality classes. \textit{J. Phys. A}
  \textbf{32}, L521 (1999).

\bibitem{GarofaloAraujoEtAl26}
J.~A. Garofalo, N.~A.~M. Ara{\'u}jo, L.~de~Arcangelis, A.~Sarracino, and
  E.~Lippiello, Janus percolation in anisotropic limited-degree networks.
  \textit{Phys. Rev. Lett.} \textbf{137}, 067401 (2026).

\bibitem{HuZiffDeng22}
H.~Hu, R.~M. Ziff, and Y.~Deng, Universal critical behavior of percolation in
  orientationally ordered {J}anus particles and other anisotropic systems.
  \textit{Phys. Rev. Lett.} \textbf{129}, 278002 (2022).

\end{thebibliography}
\end{document}